\documentclass[twocolumn,aps,showpacs,prb,superscriptaddress,floatfix,10pt,nofootinbib]{revtex4-2}
\usepackage[T1]{fontenc}
\usepackage[latin9]{inputenc}
\usepackage{amsmath}
\usepackage{amssymb}
\usepackage{esint}
\usepackage{graphicx}
\usepackage{bbold}

\usepackage{mathabx} 
\usepackage{xcolor}
\usepackage{physics}  
\usepackage{booktabs}

\definecolor{myurlcolor}{rgb}{0,0,0.7}
\definecolor{myrefcolor}{rgb}{0.8,0,0}
\usepackage[unicode=true,pdfusetitle, bookmarks=true,bookmarksnumbered=false,bookmarksopen=false, breaklinks=false,pdfborder={0 0 0},backref=false,colorlinks=true, linkcolor=myrefcolor,citecolor=myurlcolor,urlcolor=myurlcolor]{hyperref}

\newcommand{\ratio}{r} 

\makeatletter
\def\@fnsymbol#1{\ensuremath{\ifcase#1\or \dagger\or \ddagger\or
   \mathsection\or \mathparagraph\or \|\or **\or \dagger\dagger
   \or \ddagger\ddagger \else\@ctrerr\fi}}
    \makeatother

\renewcommand{\eqref}[1]{equation~(\ref{#1})}

\makeatother

\begin{document}

\title{SU(2) hadrons on a quantum computer}
\author{Yasar Atas $^*$}
\email{yyatas@uwaterloo.ca}
\affiliation{Institute for Quantum Computing, University of Waterloo, Waterloo, ON, Canada, N2L 3G1}
\affiliation{Department of Physics \& Astronomy, University of Waterloo, Waterloo, ON, Canada, N2L 3G1}
\author{Jinglei Zhang $^*$}
\email{jingleizl@gmail.com}
\affiliation{Institute for Quantum Computing, University of Waterloo, Waterloo, ON, Canada, N2L 3G1}
\affiliation{Department of Physics \& Astronomy, University of Waterloo, Waterloo, ON, Canada, N2L 3G1}
\author{Randy Lewis}
\affiliation{Department of Physics and Astronomy, York University, Toronto, ON, Canada, M3J 1P3}
\author{Amin Jahanpour}
\affiliation{Institute for Quantum Computing, University of Waterloo, Waterloo, ON, Canada, N2L 3G1}
\affiliation{Department of Physics \& Astronomy, University of Waterloo, Waterloo, ON, Canada, N2L 3G1}
\author{Jan F. Haase}
\email{jan.frhaase@gmail.com}
\affiliation{Institute for Quantum Computing, University of Waterloo, Waterloo, ON, Canada, N2L 3G1}
\affiliation{Department of Physics \& Astronomy, University of Waterloo, Waterloo, ON, Canada, N2L 3G1}
\author{Christine A. Muschik}
\affiliation{Institute for Quantum Computing, University of Waterloo, Waterloo, ON, Canada, N2L 3G1}
\affiliation{Department of Physics \& Astronomy, University of Waterloo, Waterloo, ON, Canada, N2L 3G1}
\affiliation{Perimeter Institute for Theoretical Physics, Waterloo, ON, Canada, N2L 2Y5}

\date{\today}

\begin{abstract}
	
We realize, for the first time, a non-Abelian gauge theory with both gauge and matter fields on a quantum computer. This enables the observation of hadrons and the calculation of their associated masses.
The SU(2) gauge group considered here represents an important first step towards ultimately studying quantum chromodynamics, the theory that describes the properties of protons, neutrons and other hadrons.
Quantum computers are able to create important new opportunities for ongoing essential research on gauge theories by providing simulations that are unattainable on classical computers. Our calculations on an IBM superconducting platform utilize a variational quantum eigensolver to study both meson and baryon states, hadrons which have never been seen in a non-Abelian simulation on a quantum computer.
We develop a resource-efficient approach that not only allows the implementation of a full SU(2) gauge theory on present-day quantum hardware, but further lays out the premises for future quantum simulations that will address currently unanswered questions in particle and nuclear physics.
\end{abstract}

\maketitle
\def\thefootnote{*}\footnotetext{These authors contributed equally to this work}

\section*{Introduction}

Quantum computers hold the promise to revolutionise our computational methods for understanding the fundamental interactions in Nature. Prime candidates for the application of such quantum simulations are so-called gauge theories, which play a major role in many branches of physics and comprise the entire Standard Model of particle physics. Within this area, quantum simulation of non-Abelian gauge theories is an outstanding challenge. 

The most prominent example, quantum chromodynamics (QCD), is a non-Abelian gauge theory that explains the strong interactions between quarks and gluons and ultimately underlies nuclear physics. There are also suggestions of non-Abelian forces beyond the Standard Model (BSM) that are completely separate from QCD and might, for example, underlie the Higgs sector of the Standard Model \cite{Panico:2015jxa} or provide a strongly interacting theory for dark matter \cite{Tulin:2017ara}. 

Lattice gauge theory (LGT) \cite{Wilson1974sk} is a mature and successful discretisation strategy for computational methods that has developed into an extremely successful field of science. Modern LGT calculations have provided precise quantitative results and important insights for QCD \cite{Gattringer:2010zz}, nuclear physics \cite{Detmold:2019ghl}, and non-Abelian BSM theories \cite{Kribs:2016cew}, and they will continue to do so for the foreseeable future. Quantum computers offer a possibility to extend the reach of LGT into regimes that are presently unattainable \cite{jordan_quantum_2012}. Fundamental issues like the sign problem \cite{troyer2005computational} prevent classical methods for LGTs from studying many properties of interest such as real-time particle dynamics and highly entangled matter, so quantum simulations will play an essential role in improving our understanding of Nature. 

Quantum simulations of LGTs are a growing research area \cite{Banuls:2019bmf}, addressing both real-time dynamics and equilibrium problems. Our work contributes to the latter. Equilibrium problems include important sign-problem afflicted settings such as models with high matter density (with finite chemical potentials) and topological theories. 
While current proof-of-concept demonstrations of equilibrium problems still address sign-problem free settings, they form the foundation for extensions to more complicated models. This foundation is currently built by simulating low-dimensional benchmarking models. Even though the ultimate goal is the simulation of three-dimensional (3D) theories, so far all experiments realize 1D models. Moreover, while different experimental realisations of 1D Abelian LGTs have been successful \cite{martinez_real-time_2016,klco_quantum-classical_2018,kokail_self-verifying_2019,lu_simulations_2019,mil_scalable_2020,Yang2020Observation}, non-Abelian theories are fundamentally different. Efforts to confront this challenge are underway \cite{byrnes_simulating_2006,Zohar:2012xf,Banerjee:2012xg,tagliacozzo_simulation_2013,Stannigel:2013zka,zohar_quantum_2015,silvi_finite-density_2017,rico_so3_2018,bender_digital_2018,davoudi_search_2020,raychowdhury_loop_2020,kasper_non-abelian_2020,kasper_jaynescummings_2020}, and a first important step has been made by experimentally realising pure gauge non-Abelian theories \cite{klco_su2_2020,ciavarella_trailhead_2021}. In this work we present the first quantum computer simulation for a non-Abelian gauge theory complete with dynamical matter. 

We consider as gauge group SU(2), which is the smallest non-Abelian Lie group and is thus a key step towards studying full QCD. In contrast to an Abelian theory, it is possible to build gauge singlet states from valence fermions, without any valence antifermions; the lowest energy state that exhibits this distinctly non-Abelian feature is called a baryon and it has no counterpart in an Abelian theory. The non-Abelian theory also contains a meson, which is built from one valence fermion and one valence antifermion and is thus the counterpart to a neutral state in an Abelian theory. 

While the considered non-Abelian model can in principle be realized by adapting the very successfully explored purely quantum simulations \cite{martinez_real-time_2016, Gorg2019Realization, Schweizer2019Floquet, mil_scalable_2020, Yang2020Observation}, its complexity is currently out of reach for implementing such strategies on present-day devices. We therefore use a hybrid quantum-classical approach and employ a so-called variational quantum eigensolver (VQE). Within the VQE protocols, the task of preparing the baryon and meson state is cast into the form of an optimisation problem which is solved by a classical algorithm with cost function evaluations made on a quantum co-processor.

Running deep quantum circuits on present-day devices is a formidable challenge in the current era of noisy intermediate-scale quantum (NISQ)-devices \cite{preskill_quantum_2018}, which pose severe
restrictions in the number of qubits used and the number of gates applied. Given these restrictions, we use a number of measures to make the simulations possible: (i) We integrate out the gauge field degrees of freedom to reduce the experimental resources needed. (ii) We design efficient circuits that generate an ansatz state containing only components relevant for the chosen parameter regimes and (iii) we reduce the depth of the experimental circuit by relegating part of the computation to classical preprocessing that can be performed efficiently. 

Implementing the full SU(2) theory on the IBM Quantum Experience \cite{IBM_casablanca, IBM_athens} allows us to experimentally study physics beyond the Abelian features demonstrated so far. More specifically, we experimentally access the lowest hadron energies of the model, namely the non-Abelian baryon and the meson state. This allows us to calculate their masses on the quantum computer. In particular, we perform simulations for different lattice sizes to show how a known physical symmetry emerges: the baryon and meson masses are equal in the physical limit where lattice artifacts vanish. 

\section*{Results}

\paragraph{\textbf{SU(2) gauge theory}}
The quantum field theory for SU(2) gauge fields interacting with fundamental fermions is well known  \cite{taylor_gauge_1978}. At each point in spacetime a field operator $\hat{\psi}_i$ can annihilate a fermion of color $i$ (here named red or green), or it can create the corresponding antiparticle. The gauge fields (or ``gluons'') at each point are described by a gauge link operator $\hat{U}$ which is an element of the SU(2) color group.

Because the non-Abelian nature of SU(2) leads to strong interactions and the confinement of color charge, these fermions and gluons are confined within color-singlet hadrons that cannot be studied perturbatively. In order to access the non-perturbative regime, both classical and quantum simulations require formulating the gauge theory on a lattice. Lattice calculations on classical computers are successful in Euclidean spacetime with a least-action approach, but quantum computers can address new regimes of the theory by working directly in Minkowski spacetime with a Hamiltonian approach.

\begin{figure}[t]
    \includegraphics[width=0.45\textwidth]{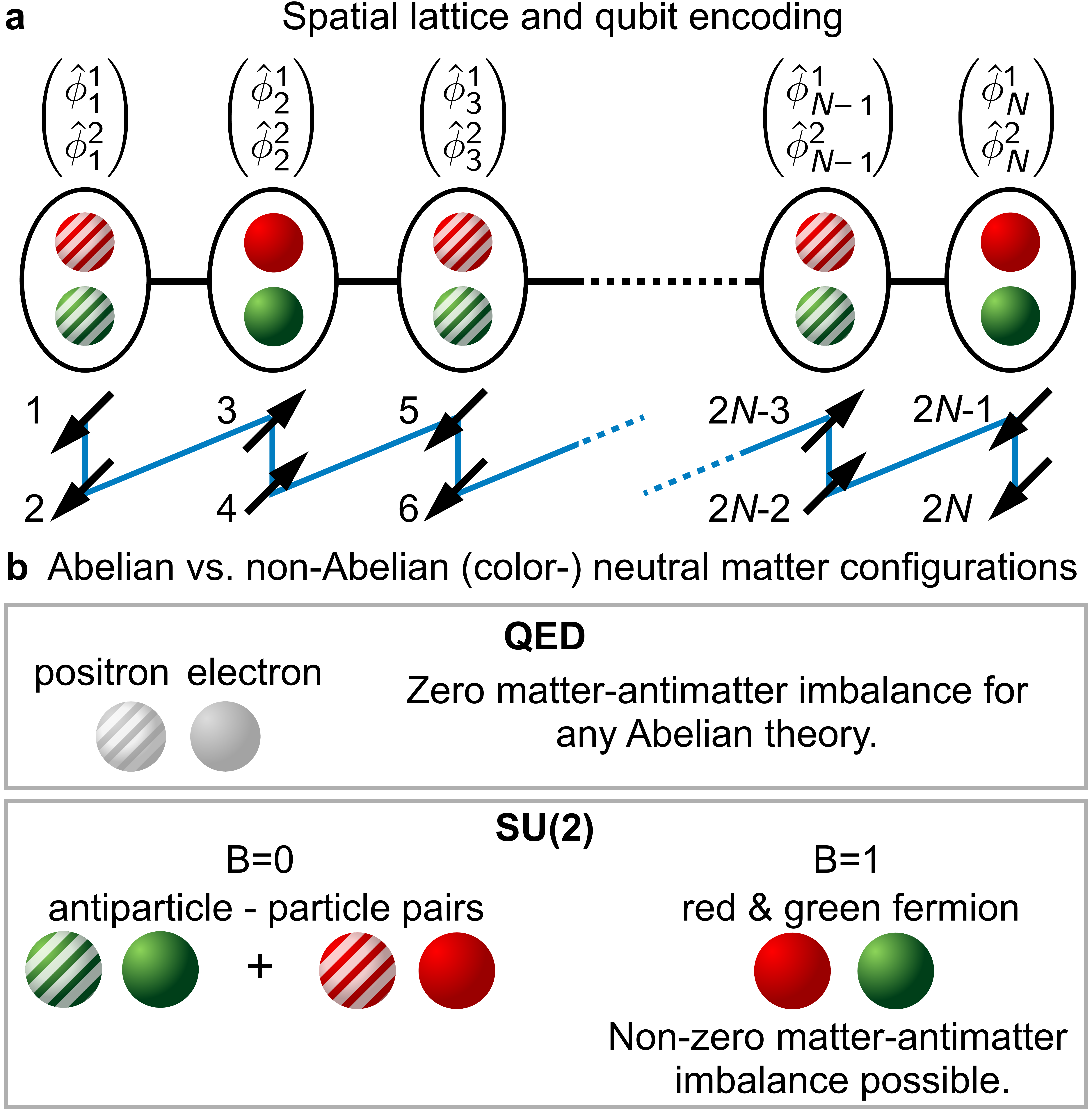} 
\caption{\textbf{Gauge theory on a lattice.} To study the SU(2) theory in one dimension, we employ the spatial lattice in panel \textbf{a}, where each site consists of either matter or antimatter particles of the the two possible colors. In the equivalent qubit formulation, each particle is represented by a qubit placed on the chain indicated by the blue line, which hence contains a number of qubits that equals twice the number of spatial sites. Panel \textbf{b} illustrates a comparison between the different gauge invariant states allowed in the neutral charge sector of Abelian QED and SU(2). While in the Abelian case neutral states require an equal number of matter (full spheres) and antimatter (striped spheres) particles, in the non-Abelian case, color-neutral states with a non-zero matter antimatter imbalance are possible.}
\label{fig:theory}
\end{figure}

We follow the staggered fermion formulation of Kogut and Susskind, where fermions and antifermions occupy separate lattice sites, arranged in an alternating pattern along the lattice (see Fig.~\ref{fig:theory}a).
The lattice Hamiltonian \cite{kogut_hamiltonian_1975,kogut_introduction_1979} in natural units ($\hbar = c = 1$) is 
\begin{equation}
	\begin{split}
		\hat{H}_l = &\frac{1}{2a_l} \sum_{n=1}^{N-1} \left( \hat{\phi}_{n}^\dagger \hat{U}_{n} \hat{\phi}_{n+1} + \operatorname{H.C.}\right)\\ & +m \sum_{n=1}^{N} (-1)^{n} \hat{\phi}_{n}^\dagger \hat{\phi}_{n}  + \frac{a_l g^{2}}{2} \sum_{n=1}^{N-1} \hat{\boldsymbol{L}}_{n}^{2},
		\label{eq:KSham}
	\end{split}
\end{equation}
where $\text{H.C.}$ denotes the Hermitian conjugate, $N$ is the number of lattice sites with spacing $a_l$, $g$ is the gauge coupling, $m$ is the fermion mass, $\hat{\phi}_n = \left( \hat{\phi}^1_n, \hat{\phi}^2_n \right)^T$ is the staggered fermion field at site $n$ with a red and a green component, and $\hat{U}_n$ is the gauge link connecting sites $n$ and $n+1$ (see Supplementary Information for the continuum model).

The last term in the Hamiltonian corresponds to the invariant Casimir operator of the theory and represents color electric field energy stored in the gauge links. Here,  $\hat{\boldsymbol{L}}_{n}^{2}=\sum_{a}\hat{L}_{n}^{a}\hat{L}_{n}^{a}=\sum_{a}\hat{R}_{n}^{a}\hat{R}_{n}^{a}$ where $\hat{L}_{n}^{a}$  and $\hat{R}_{n}^{a}$ (with $a=x,y,z$) are respectively the left and right color electric field components on the link $n$.
For a non-Abelian gauge group, the right and left color electric field are different and are related via the adjoint representation $\hat{R}_{n}^{a}=\sum_b(\hat{U}_n^{\text{adj}})_{ab}\hat{L}_{n}^{b}$, where $(\hat{U}_{n}^{\text{adj}})_{ab}=2\mathrm{Tr}\left[ \hat{U}_{n}\hat{T}^{a}\hat{U}_{n}^{\dagger}\hat{T}^{b}\right]$,  $\hat{T}^{a}=\hat{\sigma}^{a}/2$ are the three generators of the SU(2) algebra and $\hat{\sigma}^{a}$ are the Pauli matrices \cite{kogut_introduction_1979}.

\paragraph{\textbf{Symmetries and non-Abelian physics}}
By virtue of its gauge invariance, the Hamiltonian in \eqref{eq:KSham} commutes with the local gauge transformation generators, also called the Gauss's law operators, and are given by $\hat{G}_{n}^{a}\equiv \hat{L}_{n}^{a}-\hat{R}_{n-1}^{a}-\hat{Q}_{n}^{a},$ where the non-Abelian charges $\hat{Q}_{n}^{a}$ acting on the site $n$ are defined as
\begin{equation}
\hat{Q}_{n}^{a}=\sum_{ij}\hat{\phi}_{n}^{i\dagger}(\hat{T}^{a})_{ij}\hat{\phi}_{n}^{j}, \quad a=x,y,z. 
\label{nonabeliancharges}
\end{equation}   
More precisely, the so-called physical Hilbert space of the theory is spanned by the eigenstates of the Gauss's law operators $\hat{G}_{n}^{a}$. 
In the following, we choose to work in the sector with no external charges which is specified by $\hat{\boldsymbol{G}}_{n}\ket{\Psi}=0$, $\forall n$, and in the neutral total charge sector $\hat{Q}_{\text{tot}}^{a}\ket{\Psi}=\sum_{n=1}^{N}\hat{Q}_{n}^{a}\ket{\Psi}=0$, $\forall a$. 

Remarkably, the non-Abelian nature of the model allows the existence of gauge invariant singlet states which are forbidden in the Abelian case due to symmetry constraints. To see this, we note that the total color charges $\hat{Q}_{\text{tot}}^{a}=\sum_{n=1}^{N}\hat{Q}_{n}^{a}$ are conserved quantities and commute with the Hamiltonian.
Besides the three non-Abelian charges, the Hamiltonian also commutes with the redness and greenness operators defined as $\hat{\mathcal{R}} =\sum_{n=1}^{N} \hat{\phi}_{n}^{1 \dagger}\hat{\phi}_{n}^{1}-N/2$ and $\hat{\mathcal{G}} = \sum_{n=1}^{N}\hat{\phi}_{n}^{2 \dagger}\hat{\phi}_{n}^{2}-N/2$, which respectively measure the red and green color charges.
Because redness and greenness do not have convenient symmetry properties, it is more natural to use their difference (which is purely within the SU(2) gauge symmetry, since $\frac{\hat{\mathcal{R}} - \hat{\mathcal{G}}}{2} = \hat{Q}^{z}_{\text{tot}}$) and their sum (which is a global U(1) symmetry).
We therefore define the baryon quantum number of the model as 
$\hat{B} = \frac{\hat{\mathcal{R}}  + \hat{\mathcal{G}} }{2} = \frac{1}{2} \sum_{ n= 1}^{N} \hat{\phi}_{n}^{\dagger}\hat{\phi}_{n}-N/2$ which measures the matter-antimatter imbalance.

The existence of multiple conserved charges in the non-Abelian theory has to be contrasted with the Abelian $\operatorname{U}(1)$ case of quantum electrodynamics (QED), where the electric charge is the only conserved quantity. In QED, the total electric charge coincides with the baryon number $B$ of the system \cite{muschik_u1_2017}, and the neutral charge constraint thus imposes the value of the matter-antimatter imbalance to be zero. In other words, neutral gauge invariant states of QED must contain as many electrons as positrons leading to meson-type singlet states only. 
On the other hand, the constraint of neutral charge for the SU(2) theory $\hat{Q}_{\text{tot}}^{i}\ket{\Psi}=0$, $\forall i$ does not enforce the value of the baryon quantum number $B$, since these are different quantum numbers.
Therefore, it is possible to construct color neutral gauge invariant singlets with $B\neq 0$, which are forbidden in QED. While the states in the $B=0$ sector are similar to the neutral states of QED, the states in the sector with $B\neq0$ have no equivalent in Abelian theories. In particular, we will refer to the ground state in the sector with $B=1$ as a baryon state, the ground state in $B = 0$ will be the vacuum and the first excited state will be called a meson state. A pictorial comparison of a meson and a baryon is given in Fig.~\ref{fig:theory}b. 

\begin{figure}[t]
    \includegraphics[width=0.45\textwidth]{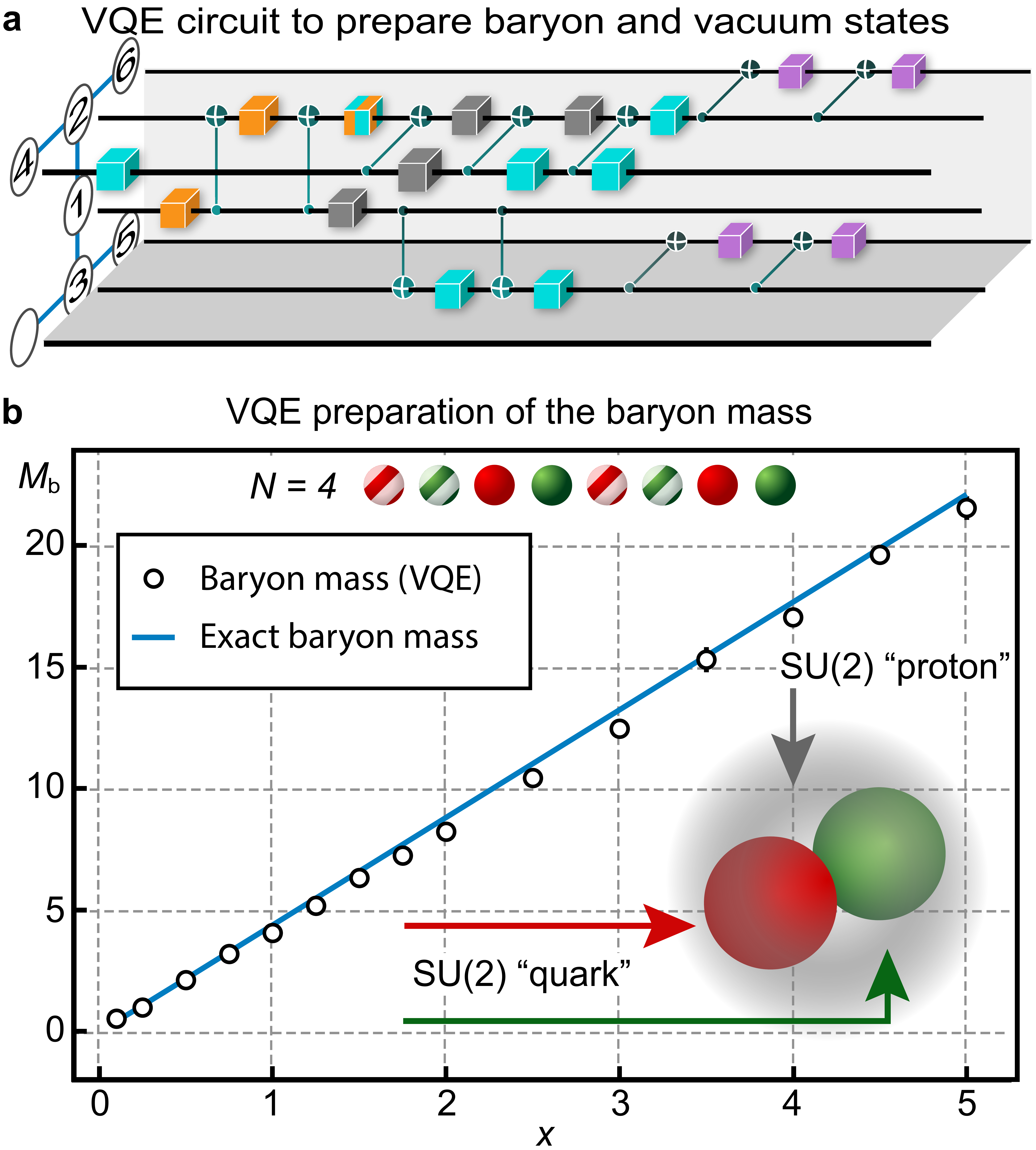} 
\caption{\textbf{VQE calculation of a baryon.} We variationally simulate an effective eight sites chain with the experimental circuit shown in \textbf{a}. The boxes represent single qubit gates. Grey boxes are fixed gates while the color coding indicates dependence from three variational parameters. Their exact implementation changes depending on the combination of the parameter values, which is automatically compiled from the original circuit shown in Fig.~\ref{fig:circuits}. This takes into account the coupling topology of the IBMQ Casablanca processor, which, together with the qubit identification for the $B=0$ sector are shown on the left. The circuit yields the mass of the baryon (errorbars are smaller than markers), an SU(2)-``proton'' (see inset), for a range of $x$ and $\tilde{m}=1$ as explained in the main text.} \label{fig:baryonVQS}
\end{figure}

\paragraph{\textbf{Elimination of the gauge fields and qubit formulation}}
To study energy spectrum of the SU(2) theory on a quantum computer, we map the lattice Hamiltonian in \eqref{eq:KSham} to a qubit system. In one spatial dimension and with open boundary conditions, the gauge degrees of freedom can be integrated out \cite{hamer_lattice_1977,ligterink_toward_2000,banuls_efficient_2017,sala_variational_2018,zohar_removing_2019} (see Supplementary Information for details).
This approach eliminates redundant degrees of freedom and allows us to simulate our target model with a minimal number of qubits. As a second step, a Jordan-Wigner transformation is applied to map the fermionic matter degrees of freedom to Pauli spin operators (see Supplementary Information for details).
The Hamiltonian is rescaled into the dimensionless form

\begin{equation}
\label{eq:hamiltonian}
	\hat{H} = \tilde{m} \hat{H}_{\text{m}} + \frac{1}{x} \hat{H}_{\text{el}} + \hat{H}_{\text{kin}},
\end{equation}
where we have defined the dimensionless Hamiltonian parameters $\tilde{m} = a_l m$, $x = \frac{1}{a_l ^{2}g^{2}}$, and we have added a constant to normalize the strong coupling ($x \rightarrow 0$) ground state energy to zero. The different terms in the Hamiltonian are given by
\begin{align}
	\hat{H}_{\text{m}} &= \sum_{n = 1}^{N} \left(\frac{(-1)^{n}}{2}\left( \hat{\sigma}_{2n-1}^{z}  + \hat{\sigma}_{2n}^{z}\right)+1 \right),\label{eq:qubitmass}\\
	\hat{H}_{\text{kin}} &= -\frac{1}{2}\sum_{n=1}^{N-1}\left( \hat{\sigma}_{2n-1}^{+}\hat{\sigma}_{2n}^{z}\hat{\sigma}_{2n+1}^{-} + \hat{\sigma}_{2n}^{+}\hat{\sigma}_{2n+1}^{z} \hat{\sigma}_{2n+2}^{-} + \text{H.C.}\right),  \label{eq:qubitkinetic}\\
\notag	\hat{H}_{\text{el}}&=\frac{3}{16}\sum_{n=1}^{N-1}(N-n)(1-\hat{\sigma}_{2n-1}^{z}\hat{\sigma}_{2n}^{z})\\  \notag &+\frac{1}{16}\sum_{n=1}^{N-2}\sum_{m>n}^{N-1}(N-m)\left(\hat{\sigma}_{2n-1}^{z}-\hat{\sigma}_{2n}^{z}\right)\left(\hat{\sigma}_{2m-1}^{z}-\hat{\sigma}_{2m}^{z}\right) \\
&+\frac{1}{2}\sum_{n=1}^{N-2}\sum_{m>n}^{N-1}(N-m)\left(\hat{\sigma}_{2n-1}^{+}\hat{\sigma}_{2n}^{-}\hat{\sigma}_{2m}^{+}\hat{\sigma}_{2m-1}^{-}+\mathrm{H.C.}\right). \label{eq:qubitelectric}
\end{align}
The Hamiltonian thus reduces to an effective qubit model with long-range interactions originating from the color electric field energy. In general, in order to simulate $N$ spatial sites, $2N$ qubits are necessary (see Fig.~\ref{fig:theory}a). The presence of non-diagonal interactions in \eqref{eq:qubitelectric} is a direct consequence of the non-Abelian character of the model (such terms are absent in the Abelian Schwinger model \cite{martinez_real-time_2016,kogut_introduction_1979,muschik_u1_2017}). Note that even though the gauge degrees of freedom no longer appear explicitly in the Hamiltonian, their interaction with the matter fields is fully taken into account. In fact, gauge field observables such as the electric field density can be computed using the reduced Hamiltonian \cite{muschik_u1_2017} and are therefore still accessible to our quantum simulation.

\begin{figure}[!ht]
    \includegraphics[width=\columnwidth]{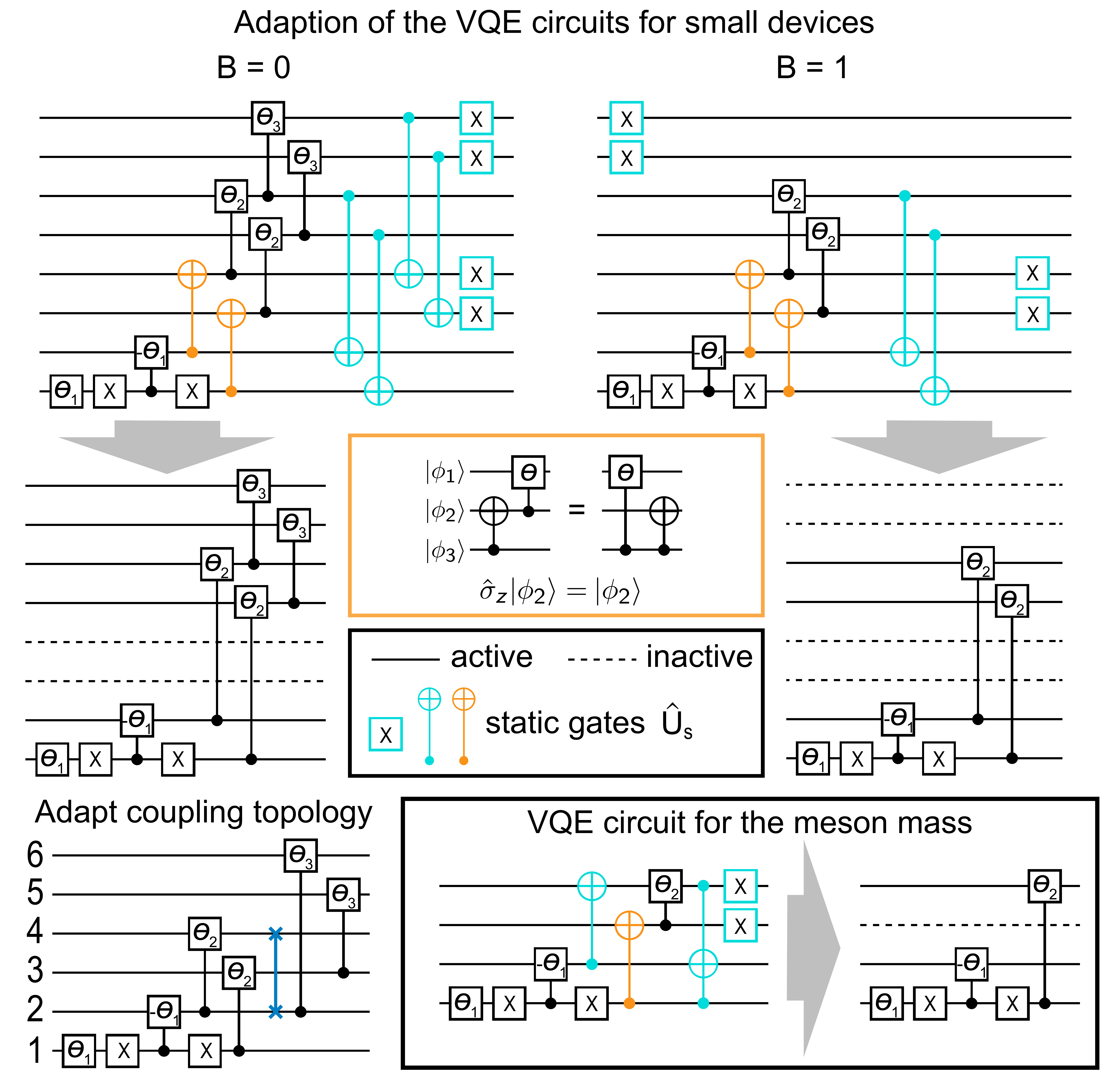} 
\caption{\textbf{VQE ansatz circuits.} The uppermost circuits for $N=4$ can be reduced by absorbing the static colored gates into $\hat{U}_s$. The parametrized controlled gates are $Y$-rotations. For the orange gates, the circuit identity in the orange box has to be applied beforehand. This results in inactive qubits (dashed lines), which do not need to be physically available on the quantum device. Details of circuit reduction are discussed in Methods. In the lower left, the introduced SWAP gate for the adaptation to the architecture of the ibmq\_casablanca processor is shown, with the qubit labeling as introduced in Fig.~\ref{fig:baryonVQS}a. The $N=2$ circuits to estimate the meson mass are illustrated in the box in the bottom right.} \label{fig:circuits}
\end{figure}
\paragraph{\textbf{Variational quantum search}}
To study the SU(2) baryon and meson states on current quantum computers, we employ the VQE approach to quantum simulations \cite{Wecker2015Progress, mcclean_theory_2016, cerezo2020variational}, which consists of a classical optimizer which aims to minimize a cost function $C(\boldsymbol{\theta})$, where $\boldsymbol{\theta}=(\theta_{1},\theta_{2},\dots)$ are the variational parameters. 
The cost function is evaluated on quantum hardware, \textit{e.g.} for the task of ground state preparation, we choose $C(\boldsymbol{\theta})=\bra{\Psi(\boldsymbol{\theta})}\hat{H}\ket{\Psi(\boldsymbol{\theta})}$ with an ansatz state $\ket{\Psi(\boldsymbol{\theta})}=\hat{U}(\boldsymbol{\theta})\ket{\Psi_{0}}$. $\ket{\Psi_{0}}$ represents a fiducial input state, and $\hat{U}(\boldsymbol{\theta})$ a parameterized unitary evolution. In our case the circuits that implement such evolution are shown in Fig.~\ref{fig:circuits}.
Our classical optimizer (see Methods) combines a mesh-based search with Bayesian optimisation techniques \cite{Frazier2018A-Tutorial}, which avoids both costly gradient estimations and convergence in a local minimum. 
For each set of parameters $\boldsymbol{\theta}$ we store the performed measurements of the multi-qubit Pauli operators contained in $\hat{H}$ (see Methods for a discussion about the decomposition of $\hat{H}$), which enables us to classically compute the corresponding value of $C(\boldsymbol{\theta})$ for different values of the Hamiltonian parameters.

To reduce the circuit depth and the number of qubits needed, \textit{i.e.} to minimize error sources on the currently available NISQ devices, we exploit the freedom to split the circuit into two parts $\hat{U}(\boldsymbol{\theta})=\hat{U}_s \hat{U}^\prime(\boldsymbol{\theta})$, where $\hat{U}_s$ contains static parts of the evolution that are not affected by the variational parameters, as shown in Fig.~\ref{fig:circuits}. Only the variational part of the circuit, $\hat{U}^\prime(\boldsymbol{\theta})$ has to be carried out on quantum hardware, as part of the computation is relegated to classical preprocessing by transforming the Hamiltonian used in $C(\boldsymbol{\theta})$ as $\hat{U}_s^{\dagger}\hat{H}\hat{U}_s$. Generally this approach comes at the cost of increasing the number of Pauli operators that have to be measured. An additional practical advantage can be gained if this decomposition results -- as in our case -- in a separable state of active and inactive qubits, $\hat{U}^\prime(\boldsymbol{\theta})\ket{\Psi_0}=\hat{u}_a(\boldsymbol{\theta})\ket{\Psi_a} \otimes \hat{u}_i \ket{\Psi_i}$ whose second component can be efficiently computed classically. Hence, only the variational part of the ansatz state $\hat{u}_a(\boldsymbol{\theta})\ket{\Psi_a}$ is implemented to measure the expectation value of the effective Hamiltonian acting on the active qubits $\bra{\Psi_i}\hat{u}_{i}^\dagger \hat{U}_s^\dagger \hat{H} \hat{U}_s \hat{u}_{i} \ket{\Psi_i}$ (see Methods for more details).

\begin{figure*}[t]
    \includegraphics[width=1\textwidth]{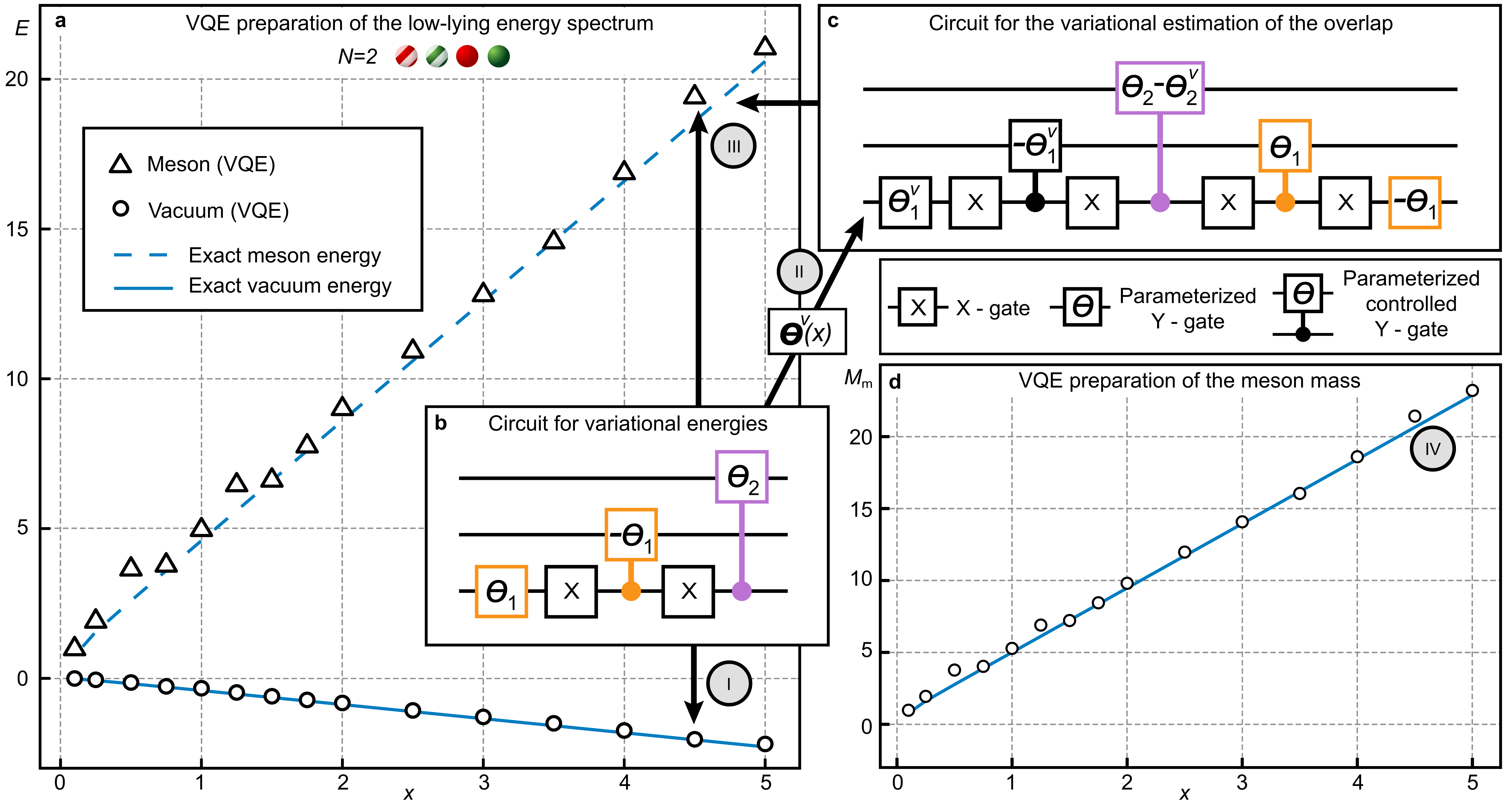} 
\caption{\textbf{VQE calculation of the meson mass.} To obtain the low-lying energy spectrum as shown in panel \textbf{a}, we first employ the circuit in \textbf{b} to obtain the vacuum energy $E_\text{v}$ (circles) in step I. Note that the employed gates are either rotations around the y-axis, the corresponding controlled gate and bitflip $X$-gates. Subsequently in step II, the variational parameters minimizing $E_\text{v}$ are used in the circuit in panel \textbf{c}, which allows to estimate the overlap betweeen the ansatz state and the variational ground state (see main text and Methods). Together with the circuit \textbf{b} we perform a VQE calculation to obtain the first excited state energy $E_{\text{m}}$ (step III, triangles). In the final step IV, we compute the energy difference $M_{\text{m}}=E_{\text{m}}-E_{\text{v}}$ and obtain the mass of the meson, shown in panel \textbf{d}. In all panels, solid or dashed lines correspond to results derived via exact diagonalisation, errorbars for experimental data are hidden due to the marker size.}\label{fig:mesonVQS}
\end{figure*}

\paragraph{\textbf{Preparation of the lightest baryon state on quantum hardware}}
Our VQE experiment determines the mass of the lightest baryon $M_{\text{b}}$, which is defined as the gap between the energy of the lowest baryon state $ E_\text{b}$ and the vacuum state $E_\text{v}$
\begin{equation}
    M_{\text{b}} = E_\text{b} - E_\text{v}.
\end{equation}
As previously discussed, the lightest baryon state is the ground state of the Hamiltonian $\hat{H}$ given in \eqref{eq:hamiltonian} in the sector with baryon number $B=1$, while the vacuum is the ground state in the sector with $B = 0$.

We experimentally prepare both states using the IBM Quantum Experience \cite{IBM_casablanca} for a lattice with $N = 4$ spatial sites, $\tilde{m} = 1$ and $x \in [0,5]$ (see Methods for a generalisation of our experimentally realized VQE scheme to larger lattices and parameter regimes). Since current quantum devices are restricted in the gate depth that can be faithfully implemented, we employ a problem-adapted efficient VQE circuit (see Fig.~\ref{fig:circuits}) that creates a limited number of basis states and variationally combines them with adjustable weights. The circuit generates only color-neutral states in the $B = 1$ symmetry sector. We further reduce the explored state space by considering only basis elements that contain up to a total of four fermions and antifermions, which approximates the ground state well in the considered parameter range.

We apply the circuit-splitting technique explained above to our ansatz state, which reduces the number of qubits from eight to four (six) for the baryon (vacuum) state, as shown in Fig.~\ref{fig:circuits}. For our case the $B=1$ baryon circuit can be seen as a specialized instance of the $B=0$ circuit, so only the more general circuit needs to be implemented on the quantum hardware, namely the lower left panel in Fig.~\ref{fig:circuits}.

The IBM Casablanca processor \cite{IBM_casablanca} consists of seven qubits with the coupling topology displayed in Fig.~\ref{fig:baryonVQS}a. We arranged the active qubits in a fashion such that only one SWAP gate is required to perform the circuit.
The reduced circuit possesses three variational parameters, each modifying several single qubit gates marked by the colored boxes in Fig.~\ref{fig:baryonVQS}a. In order to perform one measurement of the Hamiltonian we need to repeat the ansatz state preparation and measure each of the 36 multi-qubit Pauli operators in which it is decomposed, and we average the measurement results over 8024 repetitions. In order to mitigate CNOT errors this procedure is repeated three times for different noise rates, which allows to extrapolate the results to the noise-free limit (see Methods).

The baryon mass obtained from the experimental VQE is shown in Fig.~\ref{fig:baryonVQS}b and we find good agreement with the exact diagonalisation result.

\paragraph{\textbf{Accessing excited states on quantum hardware}} As a next step in studying the properties of the baryon we address its mass ratio with its partner particle, the meson. We consider the lightest meson, which is the first excited state in the $B = 0$ sector with energy $E_{\text{m}}$, and mass $M_\text{m} = E_{\text{m}} - E_{\text{v}}$.

In order to access excited states within the VQE approach, we need to modify the cost function appropriately. Since the eigenstates of the Hamiltonian are mutually orthogonal, we add a term that penalizes variational states that overlap with the lower-energy eigenstates. More precisely, after obtaining the parameters $\boldsymbol{\theta}^v$ that minimize $\bra{\Psi(\boldsymbol{\theta})}\hat{H}\ket{\Psi(\boldsymbol{\theta})}$, we consider as cost function $C(\boldsymbol{\theta}) = \bra{\Psi(\boldsymbol{\theta})}\hat{H}\ket{\Psi(\boldsymbol{\theta})} + \beta \abs{\braket{\Psi(\boldsymbol{\theta})}{\Psi(\boldsymbol{\theta}^v)}}$ to obtain the energy of the meson state, where $\beta$ is a weight chosen larger than the expected energy gap \cite{higgott_variational_2019}. The measurement of the overlap can be obtained by applying the unitary $\hat{U}(\boldsymbol{\theta})^\dagger \hat{U}(\boldsymbol{\theta}^v)$ to the initial state. This composite unitary evolution can be realized by a further application of the inverse quantum circuit. Consequently, the overlap is directly given by the probability of measuring the initial state $\ket{\Psi_0}$ in the final state $\hat{U}(\boldsymbol{\theta})^\dagger \hat{U}(\boldsymbol{\theta}^v)\ket{\Psi_0}$. This procedure is trivially extendable, \textit{i.e.} higher excited states can be obtained recursively (see Methods for more details).

Similarly to the study of the baryon, we can simplify the VQE by enforcing the suitable symmetries of the state directly within the construction of the circuit, so that it creates only basis states that have the correct $B$ number, are gauge singlets, and contain a limited number of particles. However, given the current limitations on the fidelities of available gates, calculating the required overlap is still a nontrivial task since it requires a deeper circuit. We therefore reduce our lattice size to enable the calculation on the quantum machine and simulate the properties of the meson for $N = 2$. By applying the strategies discussed in the baryon case we can reduce the number of necessary qubits from four to three.  In Fig.~\ref{fig:mesonVQS} we report the results from an experimental VQE calculation performed on the IBM Athens processor \cite{IBM_athens}, where we obtain the energies necessary to compute the meson mass. The vacuum and meson states are successfully simulated with good accuracy (Fig.~\ref{fig:mesonVQS}a), and the mass of the meson is shown in Fig.~\ref{fig:mesonVQS}d. In Fig.~\ref{fig:mesonVQS}b-c, we give the two circuits required to calculate the cost function for the simulation of the excited meson state, namely one computing the expectation value of $\hat{H}$, and one computing the overlap with the previously simulated vacuum.

\paragraph{\textbf{Path towards the continuum limit}}
\begin{figure*} 
\includegraphics[width=\textwidth]{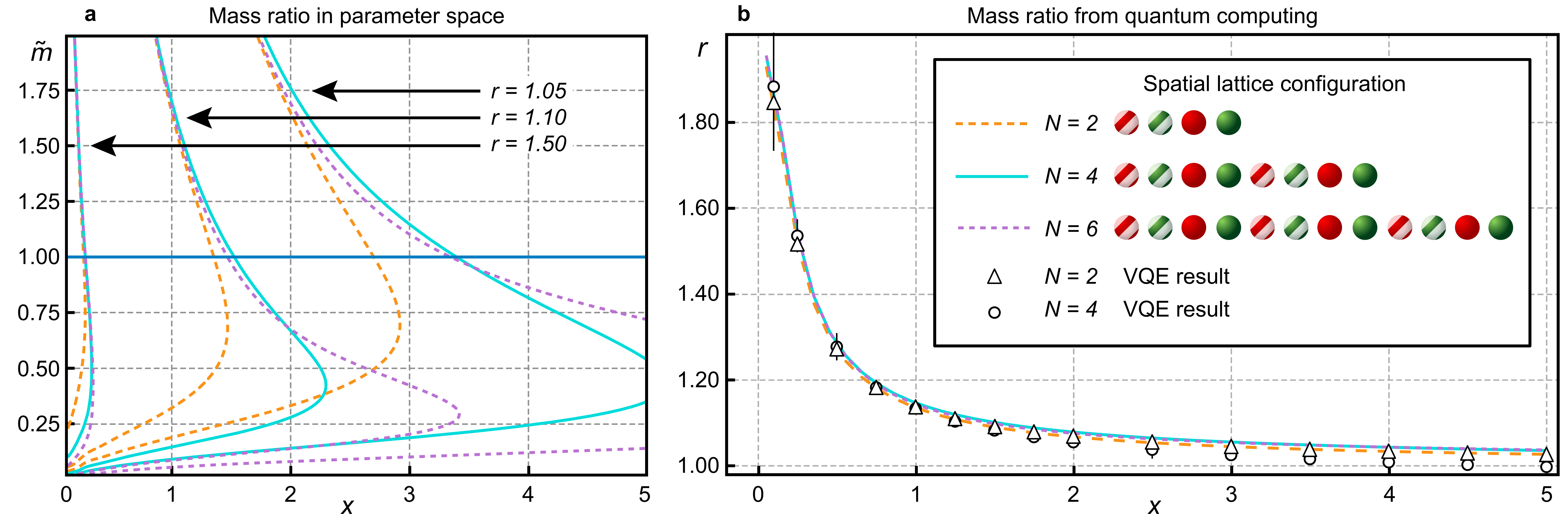} 
\caption{\textbf{Mass ratio of lightest SU(2) meson and baryon in parameter space.} Panel \textbf{a} displays lines of constant mass ratios $r$ in the $(x,\tilde{m})$ plane obtained from exact diagonalisation for lattices of size $N=2,4,6$. The blue horizontal line marks the cut shown in \textbf{b}. We supply the experimental VQE results with data obtained via exact diagonalisations, that is the energy of meson state for $N=4$ and the energy of the baryon for $N=2$. Most of the error bars are hidden by the markers.}\label{fig:ContourPlot}
\end{figure*}

In the continuum limit, SU(2) gauge theory dictates that the masses of the baryon and the meson are equal \cite{Kogut:2001na} because of a global SU(2) symmetry. Some lattice discretisations will preserve this degeneracy but for others it will only be restored in the continuum limit.
Staggered fermions \cite{kogut_hamiltonian_1975}, as used here, are in the latter category, which means the distinction between meson and baryon masses is a valuable measure of approaching the continuum limit.

To study this effect quantitatively, let us define the hadron mass ratio
\begin{equation}
    \label{eq:ratio}
    r = \frac{M_\text{m}}{M_\text{b}},
\end{equation}
and obtain this quantity with explicit calculations from the qubit Hamiltonian in equations~(\ref{eq:hamiltonian}-\ref{eq:qubitelectric}) on classical computers. 
In general, in order to extrapolate lattice calculations to the continuum limit, it is necessary to take the limit $x\to\infty$, while keeping a physical observable fixed \cite{Gattringer:2010zz}. Fig.~\ref{fig:ContourPlot}a shows curves of constant $\ratio$ in the plane spanned by $x$ and $\tilde m$, therefore we can study the continuum limit keeping a fixed ratio $\ratio$ and going to larger $x$. However, as the graph shows, any curve with a fixed $\ratio >1 $ does not allow for $x\to \infty$. The only constant-physics curve that allows it is the one in the limit of $\ratio\to 1$, therefore the correct value of the mass ratio in the continuum limit has to be $1$ as required by the theory's SU(2) global symmetry.

The large $\tilde m$ region of Fig.~\ref{fig:ContourPlot}a, \textit{i.e.} $\tilde m>1$, is also insightful.  Here the meson and baryon masses become dominated by the masses of the fermions and antifermions that they contain, relegating the meson-baryon mass difference arising from pair creation processes and gauge flux effects to be a small correction.  This is reflected in Fig.~\ref{fig:ContourPlot}a in two ways.  One is that the curves of constant $\ratio$ become independent of $x$ (getting more vertical toward the top of the graph). The other is that curves of constant $\ratio$ become independent of the lattice size $N$, since extended objects that probe the lattice boundaries always contain gauge flux which becomes a small effect at the top of the graph.

We perform experimental VQE calculations in the intermediate mass regime $\tilde m=1$ for several values of the gauge coupling within $x\in[0,5]$, marked by the horizontal blue line in Fig.~\ref{fig:ContourPlot}a.
The mass ratio $\ratio$ along this blue line is displayed in Fig.~\ref{fig:ContourPlot}b for both the experimental VQE and exact diagonalisation, and the two methods show good agreement.
The graph confirms that $r$ approaches the value $1$ for larger $x$, representing the correct restoration of mass degeneracy.

Furthermore, as is shown by comparing the exact diagonalisation data for $N = 4$ and $N = 6$, there is a clear indication that finite size effects are quite limited already for small system sizes. We show in the Methods how our experimental VQE circuit for simulating the baryon energy can be generalised to larger lattice sizes $N$, as required for the parameter range $\tilde m=1$, $x\in[0,5]$ that was discussed in Fig.~\ref{fig:ContourPlot}b. 
We also extend our study to smaller fermion masses $\tilde m$ and provide a general circuit in the Methods that allows to simulate the energy of the baryon for all parameter regimes. These extensions to our VQE experiments involve circuits beyond the capabilities of current quantum hardware and will require further experimental developments.

\section*{Discussion}

In this work, we realized the first study of a non-Abelian gauge theory including both gauge and matter fields on a quantum computer. While the gauge fields only appear implicitly in our approach, it is the gauge fields that provide the non-Abelian feature of the theory, \textit{i.e.} the existence of a gauge-invariant baryon. This proof-of-concept demonstration was made possible by a resource-efficient approach for designing our VQE circuits. While necessary to alleviate the experimental requirements for implementing the full SU(2) gauge theory, this approach also paves the way for the development of future quantum simulators.

Our work lays the foundation for a series of important next steps. Within the considered 1D SU(2)
gauge theory, our work can be extended to study other hadrons including less familiar ones such
as tetraquarks, with the goal of developing quantum simulators for nuclear physics. To this end, future
work will include the extension to SU(3) gauge theory, since that is directly applicable to
quantum chromodynamics. On this quest, extending our formulation to two and three spatial
dimensions will have to be pursued, which can be done by following \cite{haase_resource_2021}. Sign-problem afflicted
models can be considered that include for example fermionic chemical potentials or topological
terms, which both represent additions that can be included in the Hamiltonian formalism without any
fundamental roadblock.

Ultimately, LGT calculations are indispensable for studying non-Abelian gauge theories, and a dramatic new breakthrough such as quantum computing can greatly extend their range of applicability. Our simulation of a complete non-Abelian benchmarking model, including both gauge and matter fields, represents an important first step and brings a path towards the quantum computation of non-Abelian LGT into view.

\section*{Acknowledgements}
The authors wish to thank Stefan K\"uhn and Uwe-Jens Wiese for useful discussions, and Luca Dellantonio for proofreading the manuscript.
This work has been supported by Transformative Quantum Technologies Program (CFREF), NSERC, New frontiers in Research Fund, European Union's Horizon 2020 research and innovation programme under the Grant Agreement No. 731473 (FWF QuantERA via QTFLAG I03769), and US Army Research Laboratory under Cooperative Agreement Number W911NF-15-2-0060 (project SciNet).
J.F.H. acknowledges the Alexander von Humboldt Foundation in the form of a Feodor Lynen Fellowship. C.A.M. acknowledges the Alfred P. Sloan foundation for a Sloan Research Fellowship. 
We acknowledge the use of IBM Quantum services for this work. The views expressed are those of the authors, and do not reflect the official policy or position of IBM or the IBM Quantum team.

\section*{Author contributions}

J.Z., Y.A., R.L. and C.A.M. developed the theory. J.F.H., Y.A., J.Z. and R.L. developed the quantum circuits. J.F.H. developed the classical optimisation routine, and implemented the quantum simulations on the IBM hardware. A.J. performed the numerical simulations. J.Z., Y.A., J.F.H., R.L. and C.A.M. wrote the manuscript. J.F.H. led the VQE development. C.A.M. proposed and directed the work. All authors contributed to discussions of the results and the manuscript.

\clearpage

\section*{Methods}
In the following, we provide details on the methods described in the main text. In Sec.~\ref{meth:classicalPart} we discuss how the expectation value of the Hamiltonian can be measured on quantum hardware such that the VQE cost function evaluation can be performed with a minimal number of state preparations. In this section, we also describe the employed optimization algorithm. In Sec.~\ref{meth:adaptionsNISQ} we detail the strategies that we use to reduce the number of qubits and the circuit depth necessary to run the experimental VQE, while in Sec.~\ref{meth:mesonVQE} we discuss how we access excited states in our study. The specifics of the experimental implementation on IBM processors are given in Sec.~\ref{meth:implementation}. Finally, Sec.~\ref{meth:extension} presents possible VQE extensions to larger lattices and wider Hamiltonian parameter regimes.

\subsection{Hamiltonian decomposition and classical optimisation}
\label{meth:classicalPart}
The operators to be measured on the quantum hardware to estimate the expectation value $\langle \hat{H} \rangle$ can be readily obtained from equations~(\ref{eq:qubitmass}-\ref{eq:qubitelectric}) in the main text after recalling that $\hat{\sigma}^\pm = (\hat{\sigma}^x \pm i\hat{\sigma}^y)/2$. In particular, the Hamiltonian $\hat{H}$ can be written as $\hat{H}=\sum_{k=1}^n c_k(x,\tilde{m}) \hat{P}_k$, where $c_k(x,\tilde{m})$ is a real coefficient and $\hat{P}_k$ a $2N$-qubit Pauli operator, \textit{e.g.} $\bigotimes_{k=1}^{2N}\hat{\sigma}^z_k$. For our Hamiltonian, we find by direct counting that the number of Pauli strings is given by $n=6N^2-11N+9$ and grows quadratically in the number of qubits which is much smaller than the exponential upper bound of $4^{2N}$.
Hence the value of $\langle \hat{H} \rangle$ is given by $\sum_{k=1}^n c_k \langle \hat{P}_k \rangle$, where we have omitted the dependence of $c_{k}$ in $x$ and $\tilde{m}$ for simplicity. In order to reduce the number of observables to measure, we form groups of commuting operators and measure only the operator with the lowest number of identity components out of each group. This allows to calculate the expectation value of the remaining operators in the same group employing only classical computations. Note that here we restrict to local measurements of the quantum state.

During the optimisation we consider different values of the Hamiltonian parameter $x$ and it is clear that this only affects the weights $c_k(x,\tilde{m})$. Hence we can store the values of the $\langle \hat{P}_k \rangle$ obtained for different values of the variational parameters $\boldsymbol{\theta}$ and supply our optimisation routine with the updated values of $\langle \hat{H} \rangle$ after a change of $x$. 
For the estimation of the baryon mass we make use of this fact by jointly measuring all operators $\hat{P}_k^v$ and $\hat{P}_k^b$ that are required for either the vacuum ($v$) or the baryon ($b$) energy respectively. This bears two advantages: first it allows to perform the error reduction described in Methods~\ref{meth:adaptionsNISQ}, and second, it reduces the total number of calls that have to be made to the quantum processor. 

Our optimisation routine employs an intertwined combination of a grid-based search (for exploration) and a Bayesian optimizer (for exploitation) that guide each other between subsequent iterations \cite{Frazier2018A-Tutorial}. After enough refinements of the grid, it is hence guaranteed to find the global minimum. Since the optimizer accumulates more knowledge of the parameter space after each iteration and the measurements are independent of $x$, it is able to revisit any $x$-value to refine the optimisation after gaining these additional insights. On the other hand, the Bayesian techniques limit the number of optimisation parameters to around 20.

\subsection{Adaptations for NISQ hardware}
\label{meth:adaptionsNISQ}
In the following we discuss the steps to adapt the quantum circuits to the currently available hardware in more detail. 
As a concrete example, we consider the estimation of the baryon mass, where the circuits are constructed according to the targeted sector of the baryon number $B$, as outlined in the main text. Fig.~\ref{fig:circuits} in the main text focuses on the case $N=4$ and illustrates the procedure formulated in the main text. 
For both baryon numbers we can rewrite the ansatz state as $\hat{U}_s \hat{U}^\prime(\boldsymbol{\theta})\ket{\Psi_0}$, where we separate the trailing static part of the circuit which does not depend on the variational parameters and form the unitary $\hat{U}_s$. 
It becomes clear from $C(\boldsymbol{\theta})=\bra{\Psi_0}\hat{U}^{\prime\dagger}(\boldsymbol{\theta})\hat{U}_s^\dagger \hat{H} \hat{U}_s \hat{U}^\prime(\boldsymbol{\theta}) \ket{\Psi_0}$ that this corresponds to an effective transformation of the Hamiltonian $\hat{H}\mapsto \hat{U}_s^\dagger \hat{H} \hat{U}_s$ and an ansatz state produced with a shorter circuit $\hat{U}^\prime(\boldsymbol{\theta})\ket{\Psi_0}$.
Importantly, the transformation of $\hat{H}$ can be performed efficiently by applying a set of rules to the multi-qubit Pauli operators contained in it, \textit{e.g.} a CNOT with control on qubit one maps $\hat{\sigma}_x^2\hat{\sigma}_z^1 \mapsto -\hat{\sigma}_y^2\hat{\sigma}_y^1$.
This transformation does not only reduce the depth of the circuit that needs to be implemented but, crucially, also reduces the required connectivity between the qubits employed in the experiment, which usually represents a major limiting factor, especially in superconducting architectures.
Next, we note the circuit identity shown in the orange inset of Fig.~\ref{fig:circuits}, which allows to commute the two CNOT gates marked in orange with the controlled-$Y$ rotations and enables us to absorb them into $\hat{U}_s$ as well. While this identity is generally not true, here the input state $\ket{\phi_2}$ is given by $\ket{\downarrow}$, which after the application of the CNOT results in a composite Bell-like state of the type $\sqrt{p_{\downdownarrows}}\ket{\downdownarrows} + \sqrt{p_{\upuparrows}}\ket{\upuparrows}$. Hence the control qubit for the following operation can be chosen arbitrarily among them.

In a second step, we eliminate inactive qubits from the circuit. 
Note that these are not ancilla qubits in the common notion, since they are still part of the encoded quantum state of the SU(2) theory; their entanglement with other qubits has rather been traded for additional measurements that have to be performed to estimate $\langle \hat{U}_s^\dagger\hat{H} \hat{U}_s\rangle$. Nevertheless, their quantum state is now separable from the active qubits, which allows to write the ansatz as $\hat{U}^\prime(\boldsymbol{\theta})\ket{\Psi_0}=\hat{u}_a(\boldsymbol{\theta})\ket{\Psi_a} \otimes \hat{u}_i \ket{\Psi_i}$, where $\hat{u}_a(\boldsymbol{\theta})$ is the unitary containing the gates on the active qubits, while $\hat{u}_i$ corresponds to a static part that might be applied to the inactive qubits (we have $\hat{u}_i = \mathcal{I}$ for all circuits employed here). 

For example, in the case of $B=0$, the qubits three and four are now inactive, and we can therefore modify the cost function as follows, 
\begin{equation}
C(\boldsymbol{\theta}) = \bra{\Psi_a}\hat{u}_a^\dagger(\boldsymbol{\theta}) \left[\bra{\downdownarrows}_{34}\hat{U}_s^\dagger\hat{H} \hat{U}_s\ket{\downdownarrows}_{34}\right]\hat{u}_a(\boldsymbol{\theta})\ket{\Psi_a},
\end{equation}
where the term in the square brackets is an operator in the Hilbert space of the remaining qubits one, two and five to eight. 
After relabeling, we arrive at the six qubit circuit in Fig.~\ref{fig:circuits}. 
A similar procedure is performed for the circuit designed for the sector $B=1$ with the addition that also the qubits seven and eight can be removed, \textit{i.e.} the effective Hamiltonian in the brackets reads $\left[\bra{\downdownarrows\downdownarrows}_{3478}\hat{U}_s^\dagger\hat{H}\hat{U}_s\ket{\downdownarrows\downdownarrows}_{3478}\right]$, which leaves the corresponding circuit consisting of four qubits.

We remark that the six qubit circuit can be employed in both cases, if the qubits are correctly relabeled and we only remove the qubits seven and eight in the $B=1$ case. 
Since we are interested in the difference of the two eigenenergies, the latter approach is able to remove erroneous admixtures to the quantum state, since the estimations of the energies are calculated from the same sample. In more detail, for any observable $\hat{O}$ to be measured, we have $\langle \hat{O} \rangle = (1-p_e) \Tr[\hat{O} \hat{\rho}(\boldsymbol{\theta})] + p_e \Tr[\hat{O} \hat{\rho}_e]$, if we assume that at least some systematic, $\boldsymbol{\theta}$-independent part of the errors can be modeled by a convex combination to the density matrix with error probability $p_e$. 
Then $\langle \hat{O} \rangle_{B=1} - \langle \hat{O} \rangle_{B=0} = (1-p_e)\left(\Tr[\hat{O} \hat{\rho}(\boldsymbol{\theta}_{B=1})] - \Tr[\hat{O}\hat{\rho}(\boldsymbol{\theta}_{B=0})]\right)$ is independent of $\hat{\rho}_e$.

\subsection{VQE computation of the meson mass}
\label{meth:mesonVQE}
In this section we detail the experimental VQE protocol for the quantum simulation of the meson mass $E_\text{m} - E_\text{v}$. The energies $E_\text{v}$ and $E_\text{m}$ are found as the energies of the ground and first excited state in the $B=0$ subsector respectively. 
The full circuit to estimate the ground state energy for $N=2$ is shown in the lower right of Fig.~\ref{fig:circuits} in the main text. 
We reduce the circuit to three qubits by employing the methods described in Methods~\ref{meth:adaptionsNISQ}. In the main text, we explain the protocol to access the first excited state of the sector, here we cover the calculation of the overlap in more detail.
The VQE result of the ground state (\textit{i.e.} vacuum state) for a specific value of $x$ entails the parameters $\boldsymbol{\theta}^v$, such that $\hat{U}(\boldsymbol{\theta}^v)\ket{\Psi_0}$ is the ground state of $\hat{H}$ found through the quantum simulation. 
The overlap with any other state generated by a new set of variational parameters is given by $|\bra{\Psi_0}\hat{U}^\dagger(\boldsymbol{\theta})\hat{U}(\boldsymbol{\theta}^v)\ket{\Psi_0}|^2$. 
We can access the Hermitian conjugate of $\hat{U}(\boldsymbol{\theta})$ by applying the inverse circuit, \textit{i.e.} the reversed gate sequence with all parameters multiplied by $-1$. 
Since $\ket{\Psi_0}=\ket{\downarrow}^{\otimes 2N}$ is an element of the computational basis (the eigenstates of $\bigotimes_{k=1}^{2N}\hat{\sigma}^z_k$), we can obtain the required overlap as the probability of measuring the state $\hat{U}^\dagger(\boldsymbol{\theta})\hat{U}(\boldsymbol{\theta}^v)\ket{\Psi_0}$ in the computational basis and obtaining the initial state $\ket{\Psi_0}$.
Note that due to the approximately doubled circuit depth, the calculation is much more susceptible to gate errors and environmental noise processes, which can render the measurement of the overlap experimentally infeasible.

As an alternative method to obtain the first excited state, we perform a variational search in the space orthogonal to the variational ground state $\hat{U}(\boldsymbol{\theta}^v)\ket{\Psi_0}$ by implementing a Gram-Schmidt orthogonalisation procedure. The ansatz for the first excited state is thus searched in the form
\begin{equation}
\ket{\Psi_{1}(\boldsymbol{\theta})}=\mathcal{N}_{0}\left( \hat{U}(\boldsymbol{\theta})-\bra{\Psi_{0}}\hat{U}^{\dagger}(\boldsymbol{\theta}^v)\hat{U}(\boldsymbol{\theta})\ket{\Psi_0}\hat{U}(\boldsymbol{\theta}^v)\right)\ket{\Psi_0},
\end{equation}
with the normalisation factor given by
\begin{equation}
\mathcal{N}_{0}=\left(1-|\bra{\Psi_0}\hat{U}^{\dagger}(\boldsymbol{\theta}^v)\hat{U}(\boldsymbol{\theta})\ket{\Psi_0}|^{2}\right)^{-1/2}.
\end{equation} 
The new energy cost function to minimize $C(\boldsymbol{\theta}) =\bra{\Psi_{1}(\boldsymbol{\theta})} \hat{H} \ket{\Psi_{1}(\boldsymbol{\theta})}$ considers only components of the variational state orthogonal to the ground state, and explicitly reads 

\begin{equation}
C(\boldsymbol{\theta}) = \frac{\bra{\Psi(\boldsymbol{\theta})}\hat{H}\ket{\Psi(\boldsymbol{\theta})} - E_\text{v} |\bra{\Psi_0}\hat{U}^{\dagger}(\boldsymbol{\theta}^v)\hat{U}(\boldsymbol{\theta})\ket{\Psi_0}|^2}{1-|\bra{\Psi_0}\hat{U}^{\dagger}(\boldsymbol{\theta}^v)\hat{U}(\boldsymbol{\theta})\ket{\Psi_0}|^2}.
\end{equation}
This has the advantage of obtaining the excited state energy, even if the quantum circuit can only produce small components of the excited state. However, the dependence on the ground state energy $E_\text{v}$ demands a precise estimate of the latter, since the structure of the denominator implies vast distortions due to mistakes in $E_\text{v}$ when the overlap is estimated to be close to one. Hence we resort to the cost function described in the main text for the experimental calculation of the meson energy.

\subsection{Implementation on the IBM processors}
\label{meth:implementation}
All reduced circuits can be straightforwardly implemented on hardware that offers qubits arranged in a simple chain with nearest-neighbor coupling, since all further connectivity requirements have been mitigated into measurements of the effective Hamiltonian $\bra{\Psi_i}\hat{U}_s^\dagger\hat{H}\hat{U}_s\ket{\Psi_i}$ (recall $\hat{u}_i=\mathcal{I}$ here and see Methods~\ref{meth:adaptionsNISQ}).
We implement the six qubit circuit for the baryon mass on the seven qubit ibmq\_casablanca processor, which possesses the coupling map shown in Fig.~\ref{fig:baryonVQS}{b} in the main text and hence requires at least one SWAP operation. We modify the circuit as shown in Fig.~\ref{fig:circuits} in the main text and relabel the operators in the effective Hamiltonian, such that the SWAP has not to be reversed. 

Each call to the processor entails a set of calibrating circuits to mitigate readout errors, \textit{i.e.} allow one to estimate the map $\Lambda$ that mixes the true measurement probabilities $\boldsymbol{p}_{\rm{true}}$ into the observed ones $\boldsymbol{p}_{\rm{obs}} = \Lambda\, \boldsymbol{p}_{\rm{true}}$, such that one obtains an estimate of the true probabilities by an inversion of the map. 
We employ an extrapolation of the CNOT errors to mitigate their effect. Therefore we replace each CNOT in the circuits by either three or five CNOT gates to artificially enhance the effect of the introduced error and perform a linear interpolation between all results \cite{Li2017Efficient}.

\subsection{Extension for future quantum computers}
\label{meth:extension}
As explained in the previous sections, our results presented in the main text for the baryon and the meson rely on carefully chosen measures such as the mass cut-off and circuit-splitting technique. These measures are necessary given the current technological status and restrictions imposed by NISQ devices. It is however important to present tools for the investigation of our model in light of foreseeable more powerful quantum hardware. Future quantum computers will offer the possibility to use more qubits with a higher level of qubit control and increased circuit depth. To address the advantages and potential offered by these future quantum computers, we have designed circuits free of the measures taken in the case of NISQ devices. To be more precise, we first show that the VQE approach applied in our experiment can  be extended to larger lattice sizes and any parameter regime by emulating the VQE protocol on a classical device, where we estimate the mass of the baryon for $N=6$ spatial sites ($12$ qubits). The calculation involves a circuit which does not limit the number of particles that are contained in the states that are generated, but comes at the expense of a high gate depth. To alleviate the experimental requirements, we also propose an alternative circuit for the baryon in the case of $N=4$, which allows us to obtain the baryon state in all parameter regimes with high fidelity.

\subsubsection{The \texorpdfstring{$N=6$}{N=6} baryon mass}
As a proof of principle we perform numerical simulations of a VQE protocol employing a generalized ansatz circuit to estimate the baryon mass for a spatial lattice of six sites ($N=6$, 12 qubits). Here, we do not take the statistical quantum measurement noise into account.
The ansatz we choose is the following. Given a baryon number $B$, we employ the ground state at $x\to0$ (strong coupling) as our input $\ket{\Psi_0}$. For $B=1$ this corresponds to a red-green particle pair at the spatial site $N=6$ and in the case $B=0$ to the bare vacuum state. 
Note that for $B=1$, the ground state for $x\to0$ is $N/2$-fold degenerate, corresponding to the possible number of sites the red-green particle pair could occupy. For small, finite values of the parameter $x$, the kinetic term in our Hamiltonian (see \eqref{eq:qubitkinetic} in the main text) lifts the degeneracy and a second order perturbation expansion shows that the state corresponding to the red-green particle pair at $N$-th spatial site of the chain has the lowest energy, which motivates the choice of our initial state. 

\begin{figure}[t]
	\includegraphics[width=1\columnwidth]{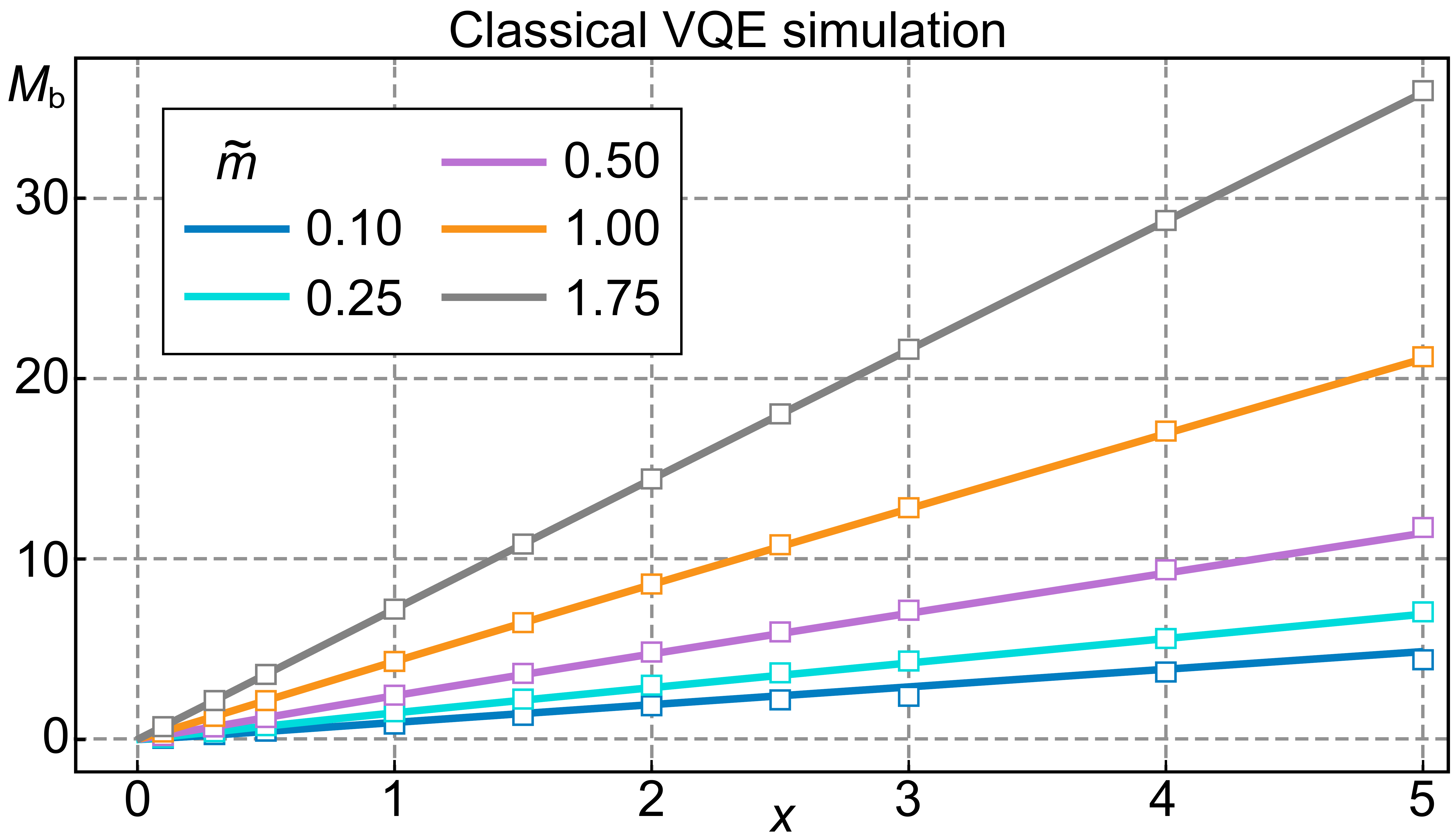} 
	\caption{\textbf{Classical simulation of a VQE to estimate the baryon mass for $N=6$.} For different values of $\tilde{m}$ we calculate the mass either via an exact diagonalisation (solid lines) or with the magnetisation preserving VQE ansatz in equation~\ref{eq:N6VQEansatz} (boxes). The case $N=4$ and $\tilde{m}=1$ calculated on real quantum hardware is shown in Fig.~2 of the main text.}
	\label{fig:N6BaryonMass}
\end{figure}

The variational circuit consists of layers of pairwise, excitation-preserving gates between neighbouring qubits, \textit{i.e.} the unitary of the $k$-th layer reads
\begin{equation}
\hat{\mathcal{U}}_k = \sum_{j=1}^{2N-1} \hat{U}_{j,j+1}(\theta_{j,k}).
\label{eq:N6VQEansatz}
\end{equation}
Here, the unitaries $\hat{U}_{j,j+1}(\theta)$ are given by parameterized SWAP gates.
Note that each $\hat{\mathcal{U}}_k$ preserves the total spin $\langle \hat{\sigma}_{\text{tot}}^z\rangle$ and hence ensures that the final state also lies in the chosen subspace characterized by $B$. In fact, it can be easily shown that the baryon number in qubit formulation is given by $ \hat{B} = \frac{\hat{\sigma}^z_{\text{tot}}}{4}$, and therefore subspaces with fixed baryon number correspond to subspaces with fixed total magnetisation.
Employing 10 (15) layers in the $B=0$ ($B=1$) sector, we obtain the baryon mass shown in Fig.~\ref{fig:N6BaryonMass} for different values of $\tilde{m}$. Importantly, this procedure grants access to the whole parameter space illustrated in Fig.~\ref{fig:ContourPlot} of the main text.
\newline

\begin{figure*}[t]
	\includegraphics[width=1\textwidth]{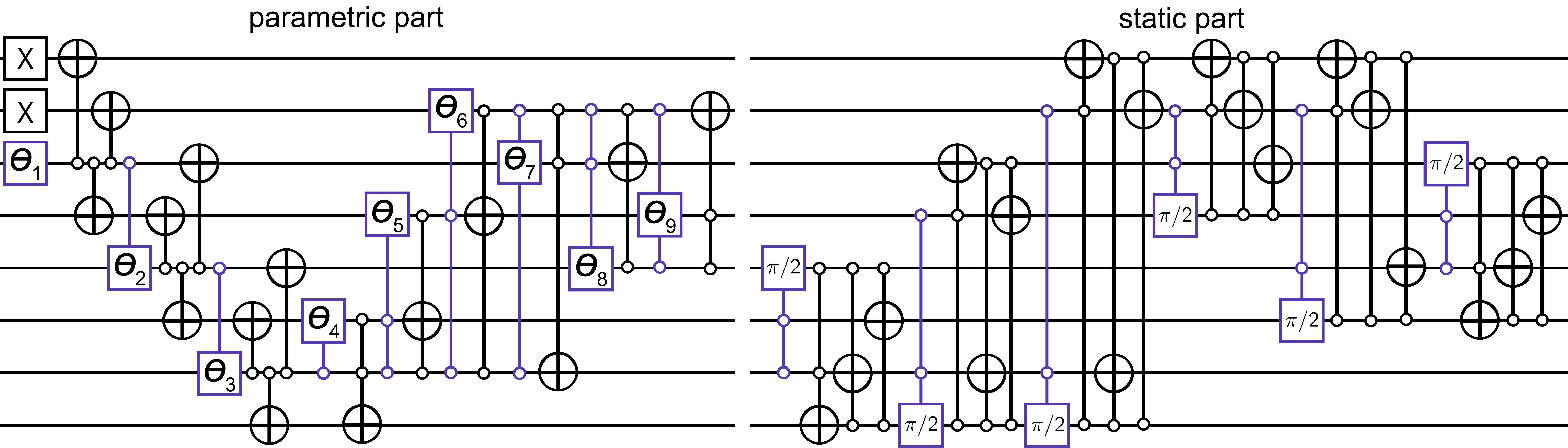} 
	\caption{\textbf{General baryon circuit for $N=4$} The parametric part of the circuit involves nine variational parameters, while the static part can be incorporated into the Hamiltonian to reduce the computational effort as discussed in the main text. The colored gates mark (controlled) rotations around the $Y$-axis. White control marks denote active application of the gate when the control is in $\ket{\downarrow}$.  The circuit, when applied to the initial state $\ket{\upuparrows \upuparrows \upuparrows \upuparrows}$, generates the $16$ basis states satisfying the $B=1$ symmetry (reported in Table \ref{t1}) and combines them to form color singlet thus reducing the total number of necessary variational parameters. Classical simulations of noiseless VQE using this circuit have demonstrated a high fidelity with the exact ground state in the $B=1$ sector.}
	\label{SIfigure2}
\end{figure*}

\subsubsection{General circuit for the \texorpdfstring{$N=4$}{N=4} baryon}
In order to further lower the computational effort of the previous brute force approach, we propose another solution that reduces the depth of the circuit and number of variational parameters. We consider a circuit that generates the ansatz state for the lightest baryon on a lattice with $N=4$ sites and arbitrary Hamiltonian parameters. 

In table \ref{t1}, we have listed all the basis states whose total magnetisation is equal to $4$ (corresponding to a baryon quantum number $B=1$), and are annihilated by $\hat{Q}_{\mathrm{tot}}^{z}$. There are $16$ such states and $12$ of them must be combined pairwise in order to form color singlet combinations, \textit{i.e} such that an application of the three non-Abelian charges $\hat{Q}_{\mathrm{tot}}^{a}$ with $a=x,y,z$ is equal to zero. Basis states which have to be combined are written with the same numeric index and located in the same row of Table~\ref{t1}. For instance, the two states appearing in the fifth row must be combined in the following way 
\begin{equation}
\ket{\tilde{s}_{5}}=\frac{1}{\sqrt{2}}\left( \ket{s_{5}}-\ket{s_{5}^{\prime}}\right)
\label{color_singlet_example}
\end{equation}
to be a common eigenstate of the three non-Abelian charges with eigenvalue zero (for more detail, see Supplementary Information).
For larger lattice sizes $N$, the construction of the singlet states becomes more involved and will not be addressed here.

\begin{table}[]
	\centering
	\begin{tabular}{c}
		\toprule
		$N = 4, B = 1$ basis states \\
		\midrule
		$\ket{s_1} = \ket{\upuparrows \upuparrows \upuparrows \downdownarrows}$ \\
		\midrule
		$\ket{s_2} = \ket{\upuparrows \upuparrows \downdownarrows \upuparrows}$\\
		\midrule
		$\ket{s_3} = \ket{\upuparrows \downdownarrows \upuparrows \upuparrows}$\\
		\midrule
		$\ket{s_4} = \ket{\downdownarrows \upuparrows \upuparrows \upuparrows}$\\   
		\midrule
		$\ket{s_5} = \ket{\updownarrows \downuparrows \upuparrows \upuparrows}$ \\
		$\ket{s_5^\prime} = \ket{\downuparrows \updownarrows \upuparrows \upuparrows}$\\ 
		\midrule
		$\ket{s_6} = \ket{\updownarrows  \upuparrows \downuparrows \upuparrows}$ \\
		$\ket{s_6^\prime} = \ket{\downuparrows \upuparrows \updownarrows  \upuparrows}$\\ 
		\midrule
		$\ket{s_7} = \ket{\updownarrows  \upuparrows \upuparrows \downuparrows }$ \\
		$\ket{s_7^\prime} = \ket{\downuparrows \upuparrows  \upuparrows \updownarrows }$\\ 
		\midrule
		$\ket{s_8} = \ket{ \upuparrows \upuparrows\updownarrows  \downuparrows }$ \\
		$\ket{s_8^\prime} = \ket{ \upuparrows  \upuparrows \downuparrows \updownarrows }$\\ 
		\midrule
		$\ket{s_9} = \ket{ \upuparrows \updownarrows \upuparrows \downuparrows }$ \\
		$\ket{s_9^\prime} = \ket{ \upuparrows \downuparrows \upuparrows  \updownarrows }$\\ 
		\midrule
		$\ket{s_{10}} = \ket{ \upuparrows \updownarrows \downuparrows  \upuparrows}$ \\
		$\ket{s_{10}^\prime} = \ket{ \upuparrows \downuparrows   \updownarrows \upuparrows}$\\ 
		\bottomrule
	\end{tabular}
	\caption{All basis states with baryon number $B=1$ for a lattice with $N=4$ spatial sites. States in the same row must be combined together to form a color singlet as exemplified in \eqref{color_singlet_example}.}
	\label{t1}
\end{table}

Once we have constructed a basis for the $B=1$ symmetry sector composed of color singlet states, we can parametrize an ansatz for the lightest baryon considering a superposition of such basis elements with real coefficients. For example we can use hyperspherical coordinates and consider
\begin{equation}
\ket{\Psi(\boldsymbol{\theta})}=\sum_{n=1}^{10} a_{n}(\boldsymbol\theta)\ket{\tilde{s}_{n}}, \label{eq:ansatzstate}
\end{equation}
where $\ket{\tilde{s}_{n}}=(\ket{s_{n}}-\ket{s_{n}^{\prime}})/\sqrt{2}$ are the color singlet combinations of basis states appearing in table \ref{t1}, $a_{n}(\boldsymbol{\theta})=\prod_{i=1}^{n-1}\sin (\theta_{i})\cos(\theta_{n})$ for $n=1,2,\dots,9$, $a_{10}(\boldsymbol{\theta})=\prod_{i=1}^{9}\sin(\theta_{i})$, and $\boldsymbol{\theta}=(\theta_{1},\theta_{2}\dots \theta_{9})$ is a vector of nine variational parameters. Note that only nine parameters are required to describe the ansatz state since the tenth is automatically fixed by the normalisation. The circuit generating the ansatz state is represented in Fig.~\ref{SIfigure2} and has been separated into two parts. The first parametric part contains the nine variational parameters and creates the following superposition
\begin{equation}
\ket{\psi(\boldsymbol{\theta})}=\hat{U}^\prime(\boldsymbol{\theta})\ket{\upuparrows \upuparrows \upuparrows \upuparrows}=\sum_{n=1}^{10}a_{n}(\boldsymbol{\theta})\ket{s_{n}}
\end{equation}
where $\hat{U}^\prime(\boldsymbol{\theta})$ is the unitary representing the parametric part of the circuit, $\ket{\upuparrows \upuparrows \upuparrows \upuparrows}$ is the input state and $\ket{s_{n}}$ are the basis states in table \ref{t1}. The purpose of the second part of the circuit is to impose the color symmetry and hence to produce the color symmetric superpositions as in \eqref{eq:ansatzstate}, \textit{i.e} $\ket{\Psi(\boldsymbol{\theta})}=\hat{U}_{s}\ket{\psi(\boldsymbol{\theta})}=\hat{U}_{s}\hat{U}^\prime(\boldsymbol{\theta})\ket{\upuparrows \upuparrows \upuparrows \upuparrows}$ with $\hat{U}_{s}$ the unitary representing the static part of the circuit in Fig.~\ref{SIfigure2}. 
Let us note that the static part of the circuit possesses a block structure with an elementary block made of a double controlled $\pi/2$ $Y$-rotation followed by three Toffoli gates. There are six elementary blocks corresponding to the six states which need to be combined in Tab.~\ref{t1}. As an example, let us consider the first block and see its action on the state $\ket{\psi(\boldsymbol{\theta})}$. The controlled rotation acts only on the basis state $\ket{s_{5}}$, which is the only one having both spins pointing down at positions two and three, and hence generates the state $\ket{\updownarrows \downdownarrows \upuparrows \upuparrows }$ with weight $-a_{5}(\boldsymbol{\theta})$. The three subsequent Toffoli gates transform this state into $\ket{s_{5}^{\prime}}$, resulting in the color singlet state given in \eqref{color_singlet_example}. The other blocks act similarly, and after application of the static part, we obtain the ansatz state given in \eqref{eq:ansatzstate}. 
Also note that the overall structure of the circuit would in principle allow the use of the splitting technique to further reduce the computational effort as described in the previous sections.
Classical simulations of noise free VQE with this circuit have demonstrated high fidelity with the exact ground state in the $B=1$ sector and for any value of the Hamiltonian parameters.

\clearpage

\section*{Supplementary Information}
The purpose of this Supplementary Information is to provide details about the derivation of the results presented in the main text and methods. In Sec.~\ref{SI:continuum}, we introduce the continuum non-Abelian SU(2) Yang-Mills Hamiltonian and show how its discretisation leads to the lattice Kogut-Susskind Hamiltonian of the main text. In Sec.~\ref{SI:elimination}, we explain the unitary transformation which allows us to eliminate the gauge fields from the Hamiltonian. In particular, we show how Gauss's laws take a simple form in a rotated frame and can be easily integrated leading to long-range interactions between non-Abelian charges in the color electric field Hamiltonian. In Sec.~\ref{SI:qubit}, we perform a Jordan-Wigner mapping in order to express the Hamiltonian in terms of qubits degrees of freedom, in order to be able to implement the system on quantum hardware.  
Finally, in Sec.~\ref{SI:singletstates} we discuss the construction of a color symmetric basis that we use in order to design the circuit for the ansatz state of our VQE experiment. 

\subsection{SU(2) gauge theory in the continuum}
\label{SI:continuum}
In the following, we consider the continuum version of the SU(2) Yang-Mills Hamiltonian and show how its discretisation leads to the Kogut-Susskind Hamiltonian given in \eqref{eq:KSham} of the main text. 
We regard the non-Abelian SU(2) Yang-Mills model describing fermions and antifermions interacting via color electric fields in one spatial dimension. In this model, the fermions and antifermions carry a color charge which we will refer to as red and green. The interactions between the fermions are mediated by the gauge field or ``color electric field''. We denote the gauge field vector potential at position $z$ with temporal and spatial components $\hat{A}_{0}^{a}(z)\hat{T}^{a}$ and $\hat{A}_{1}^{a}(z)\hat{T}^{a}$ respectively, where $\hat{T}^{a}=\hat{\sigma}^{a}/2$ are the three generators of the $\operatorname{SU}(2)$ Lie algebra, and $\hat{\sigma}^{a}$ the $a$-th Pauli matrix ($a=x,y,z$). From now on, we adopt the temporal or Weyl gauge $\hat{A}_{0}^{a}(z)=0$ and Einstein's summation convention will apply on repeating indices in color space, but not on lattice indices. The color electric field also carries a group index $a$ and is given by $\hat{L}^{a}(z)=-\partial_{t}\hat{A}_{1}^{a}(z)$. As the canonical conjugate momentum of $\hat{A}_{1}^{a}(z)$, the color electric field satisfies $[\hat{A}_{1}^{a}(z),\hat{L}^{b}(y)]=i\delta_{ab}\delta(z-y)$. In the continuum, the Yang-Mills Hamiltonian is given by \cite{jackiw1980introduction}
\begin{equation}
\begin{split}
\hat{H}_{\mathrm{cont}} =\int \mathrm{d}z  \, & \left[ {\vphantom{\frac{1}{2}}} \hat{\bar{\psi}}(z)\gamma^{1}\left( -i\partial_{z}+g\hat{A}_{1}^{a}(z)\hat{T}^{a}\right)\hat{\psi}(z) \right. \\
& \left. + m \hat{\bar{\psi}}(z)\hat{\psi}(z)+\frac{1}{2}\hat{L}^{a}(z)\hat{L}^{a}(z) \right],
\end{split} \label{SIcontinuum_Ham}
\end{equation}
where $\hat{\psi}(z)=(\hat{\psi}_{1}(z),\hat{\psi}_{2}(z))^{T}$ is a two-component spinor representing the matter fields. The fermion mass is denoted by $m$, and $g$ quantifies the matter-field coupling constant, while $\gamma^{\mu}$ are the Dirac matrices satisfying the anticommutation relations $\lbrace \gamma^{\mu},\gamma^{\nu} \rbrace =2\eta^{\mu \nu}$ with $\eta=\mathrm{diag}(1,-1)$ the metric tensor. We further use the shorthand notation $\hat{\bar{\psi}}(z)=\hat{\psi}^{\dagger}(z)\gamma^{0}$. 
In one dimension, a convenient representation for the Dirac matrices is given by $\gamma^{0}=\sigma^{z}$ and $\gamma^{1}=i\sigma^{y}$. Note that the first term in equation~(\ref{SIcontinuum_Ham}) represents the gauge-invariant kinetic energy, the second contribution corresponds to the mass term, and the last part gives the color electric energy.

For quantum (and classical) simulation purposes, it is more convenient to work with a discretised version of the continuum Hamiltonian, which is defined on a spatial lattice whose points are separated by a distance $a_l$. In this work, we adopt the staggered formulation of Kogut and Susskind, where fermions and antifermions occupy separate lattice sites, and are arranged in an alternating pattern along the lattice. This yields the lattice Hamiltonian
\begin{equation}
\begin{split}
\hat{H_l} = &\frac{1}{2a_l} \sum_{n=1}^{N-1} \left( \hat{\phi}_{n}^\dagger \hat{U}_{n} \hat{\phi}_{n+1} + \operatorname{H.C.}\right)\\ & +m \sum_{n=1}^{N} (-1)^{n} \hat{\phi}_{n}^\dagger \hat{\phi}_{n}  + \frac{a_l g^{2}}{2} \sum_{n=1}^{N-1} \hat{\boldsymbol{L}}_{n}^{2}. \label{SIeq:KSham}
\end{split}
\end{equation}

The matter field at each lattice site $n$ is described by a two-component fermionic field $\hat{\phi}_{n}=(\hat{\phi}_{n}^{1},\hat{\phi}_{n}^{2})^{T},$
where the upper index labels the two possible colors. These are related to the components of the continuum spinor in equation~(\ref{SIcontinuum_Ham}) in the limit of vanishing lattice spacing $a_l \rightarrow 0$ as $i^{p}(-1)^{n}\hat{\phi}_{2n+p}/\sqrt{a_l}\rightarrow \hat{\psi}_{p+1}(z)$ with $p=0,1$.
In the first (kinetic) term, the parallel transporter (or connection) $\hat{U}_n=\exp(i\hat{\Omega}_{n}^{a}\hat{T}^{a})$ acts on the link between sites $n$ and $n+1$ and mediates the interaction between the internal color degree of freedom of the fermions on neighbouring sites. Its presence ensures the invariance of the Hamiltonian under local gauge transformations. The angular variables $\hat{\Omega}_{n}^{a}$ are related to the continuum gauge field on the link  $n$ as $\hat{\Omega}_{n}^{a}/(a_l g) \rightarrow \hat{A}_{1}^{a}(z)$ when the lattice spacing goes to zero.

The last term in the Hamiltonian corresponds to the invariant Casimir operator of the theory and represents the color electric field energy stored in the gauge links. More precisely,  $\hat{\boldsymbol{L}}_{n}^{2}=\hat{L}_{n}^{a}\hat{L}_{n}^{a}=\hat{R}_{n}^{a}\hat{R}_{n}^{a}$ where $\hat{L}_{n}^{a}$  and $\hat{R}_{n}^{a}$ (with $a=x,y,z$) are respectively the left and right color electric field components on link $n$. They are conjugate momenta of the vector potential \cite{kogut_introduction_1979} and are related to the continuum variable via $\hat{L}_{n}^{a}\rightarrow \hat{L}^{a}(z)/g$. The operators $\hat{L}_{n}^{a}$ and $\hat{R}_{n}^{a}$ satisfy the algebra $[\hat{R}_{n}^{a},\hat{R}_{m}^{b}]=i\epsilon_{abc}\hat{R}_{n}^{c}\delta_{mn}$,  $[\hat{L}_{n}^{a},\hat{L}_{m}^{b}]=-i\epsilon_{abc}\hat{L}_{n}^{c}\delta_{mn}$, and $[\hat{L}_{n}^{a},\hat{R}_{m}^{b}]=0$, where $\epsilon_{abc}$ is the Levi-Civita symbol.
For a non-Abelian gauge group, the right and left color electric field are related via the adjoint representation $\hat{R}_{n}^{a}=(\hat{U}_n^\text{adj})_{ab}\hat{L}_{n}^{b}$, with 
\begin{equation}
(\hat{U}_{n}^\text{adj})_{ab}=2\mathrm{Tr}\left[ \hat{U}_{n}\hat{T}^{a}\hat{U}_{n}^{\dagger}\hat{T}^{b}\right].
\label{adjoint_def}
\end{equation}
Concluding, if one follows the mapping between the lattice and continuum variables prescribed above, and employs the correspondence
$a_l \sum_{n=1}^{N}f(a_l n)\rightarrow \int \mathrm{d}z\, f(z)$ as $a_l \rightarrow 0$, one can recover the continuum Hamiltonian in equation~(\ref{SIcontinuum_Ham}) from the lattice version given in equation~(\ref{SIeq:KSham}).

\subsection{Elimination of the gauge fields}
\label{SI:elimination}
For one spatial dimension and open boundary conditions the gauge fields can be eliminated and expressed in terms of the fermionic fields \cite{hamer_lattice_1977,sala_variational_2018}. The decoupling is achieved by a unitary transformation $\hat{{\Theta}}$ that acts on the fermionic fields and eliminates the gauge connections $\hat{U}_n$ from the kinetic term and expresses the color electric energy in terms of the fermionic operators. This approach was recently introduced by the authors of \cite{sala_variational_2018}, who performed tensor-network simulations of the SU(2) gauge theory to study real time dynamics of string breaking phenomena.
In the following, we reproduce the main steps of their derivation.

We seek a unitary transformation $\hat{\Theta}$ such that $\hat{\Theta} \left(\hat{\phi}_{n}^{\dagger}\hat{U}_{n}\hat{\phi}_{n+1}\right) \hat{\Theta}^{\dagger} =\hat{\phi}_{n}^{{\dagger}}\hat{\phi}_{n+1}$. Since the $\hat{U}_{n}$ are unitary, one may demand the action of $\hat{\Theta}$ on the fermionic field to be $\hat{\Theta} \hat{\phi}_{n}\hat{\Theta}^{\dagger} =\hat{U}_{n-1}^{\dagger}\hat{U}_{n-2}^{\dagger}\cdots \hat{U}_{1}^{\dagger} \hat{\phi}_{n}$.
One hence introduces the operators
\begin{equation}
\hat{W}_{k}=\exp \left(i\, \hat{\boldsymbol{{\Omega}}}_{k}\cdot\sum_{m>k} \hat{\boldsymbol{Q}}_{m}\right),
\end{equation}
where $\hat{\boldsymbol{Q}}_{m}$ is the vector of the non-Abelian charges with components
\begin{equation}
\hat{Q}_{m}^{a}=\hat{\phi}_{m}^{i\dagger}(\hat{T}^{a})_{ij}\hat{\phi}_{m}^{j}, \quad a=x,y,z. \label{si:nonabeliancharges}
\end{equation}
The three (operator) components vector $\hat{\boldsymbol{{\Omega}}}_{k}= -a_l g\boldsymbol{\hat{A}}_{1}$ is directly proportional to the spatial component of the gauge field at site $k$, and are related to the parallel transporters, since $\hat{U}_{k}=\exp\left( i \, \hat{\Omega}_{k}^{a}\hat{T}^{a} \right)$. In \cite{sala_variational_2018}, it was shown that the desired $\hat{\Theta}$-transformation is explicitly given by $\hat{\Theta}=\hat{W}_{1}\hat{W}_{2}\cdots \hat{W}_{N}$. Under this transformation, the Kogut-Susskind Hamiltonian given by equation~(2)  takes the form
\begin{align}
\notag \hat{H}_{\text{rot}}\equiv\hat{\Theta} \hat{H} \hat{\Theta}^{\dagger}=&\frac{1}{2a_l} \sum_{n=1}^{N-1}\left( \hat{\phi}_{n}^{{\dagger}}\hat{\phi}_{n+1}+\mathrm{H.C.}\right) \\
& +m\sum_{n=1}^{N}(-1)^{n}\hat{\phi}_{n}^{\dagger}\hat{\phi}_{n}+\frac{a_l g^{2}}{2} \hat{H}_{e}, \label{SIeqrotatedHam}
\end{align}
where the rotated color electric term is given by
\begin{equation}
\hat{H}_{e}\equiv \sum_{n=1}^{N-1}\hat{\Theta} \hat{\boldsymbol{L}}_{n}^{2} \hat{\Theta}^{\dagger}= \sum_{n=1}^{N-1}\left(\hat{\boldsymbol{R}}_{0}+\sum_{m\leq n}\hat{\boldsymbol{Q}}_{m}\right)^{2}, \label{SIeq:rotatedelec}
\end{equation}
and $\hat{\boldsymbol{R}}_{0}$ can now be interpreted as a background field.

We now give more details on how Gauss's law has been used in order to arrive at equation (\ref{SIeq:rotatedelec}).
In the initial frame (before the $\hat{\Theta}$ transformation),  Gauss's law reads
\begin{equation}
\hat{G}_{n}^{a}\equiv \hat{L}_{n}^{a}-\hat{R}_{n-1}^{a}-\hat{Q}_{n}^{a}, \quad a=x,y,z,
\end{equation}
where $\hat{L}_{n}^{a}$ and $\hat{R}_{n-1}^{a}$ act on the links emanating from the site $n$, which itself carries the non-Abelian color charge $\hat{Q}_{n}^{a}$. Gauss's law operators $\hat{G}_{n}^{a}$ are the generators of the local gauge transformations, hence we have $[\hat{H},\hat{G}_{n}^{a}]=0, \forall n, \forall a$. Since we assume that there are no external charges in the system, the gauge invariant states must satisfy the identity $\hat{L}_{n}^{a} - \hat{R}_{n-1}^{a} = \hat{Q}_{n}^{a}$. 

In order to see how Gauss's law transforms under $\hat{\Theta}$, one needs to find the transformation rules for $\hat{L}_{n}^{a}$, $\hat{Q}_{n}^{a}$, and $\hat{R}_{n-1}^{a}$. Using the following commutation relation \cite{kogut_hamiltonian_1975}
\begin{equation}
[ \hat{L}_{n}^{a},(\hat{U}_{n})_{pq}]=(\hat{T}^{a}\hat{U}_{n})_{pq} \label{commutation_L_U},
\end{equation}
and recognising that $\hat{W}_{n}$ has the same matrix structure as $\hat{U}_{n}$, we find \cite{sala_variational_2018}
\begin{equation}
[ \hat{L}_{n}^{a},\hat{W}_{n}]=\left( \sum_{m>n}\hat{Q}_{m}^{a}\right)\hat{W}_{n},
\end{equation}
which leads to 
\begin{equation}
\hat{W}_{n}\hat{L}_{n}^{a}\hat{W}_{n}^{\dagger}=\hat{L}_{n}^{a}-\sum_{m>n}\hat{Q}_{m}^{a}.
\end{equation}
As a consequence, under the $\hat{\Theta}$ transformation the left electric field transforms as 
\begin{equation}
\hat{\Theta}\hat{L}_{n}^{a}\hat{\Theta}^{\dagger}=\hat{L}_{n}^{a}-\sum_{m>n}\hat{W}_{1}\cdots \hat{W}_{n-1}\hat{Q}_{m}^{a}\hat{W}_{n-1}^{\dagger}\cdots \hat{W}_{1}^{\dagger}.
\end{equation}
To find how the non-Abelian charges transform under $\hat{W}_{k}$, we make use of the following identity 
\begin{equation}
\hat{W}_{k}\hat{\phi}_{n}^{i}\hat{W}_{k}^{\dagger}=(\hat{U}_{k})_{ij}\hat{\phi}_{n}^{j},
\end{equation}
which can be easily derived from the commutation relation $[\hat{Q}_{n}^{a},\hat{\phi}_{m}^{i}]=-\delta_{mn}(\hat{T}^{a})_{ij}\hat{\phi}_{n}^{j}$. By using the definition of the non-Abelian charges in terms of the fermionic field given in equation (\ref{si:nonabeliancharges}), we find 
\begin{equation}
\hat{W}_{k}\hat{Q}_{m}^{a}\hat{W}_{k}^{\dagger}=\hat{\phi}_{m}^{\dagger i}(\hat{U}_{k})_{i p} (\hat{T}^{a})_{pq}(\hat{U}^{\dagger}_{k})_{q j}\hat{\phi}_{m}^{j}=(\hat{U}_{k}^{\text{adj}})_{ab}\hat{Q}_{m}^{b},
\label{QtransformationbyW}
\end{equation}
where the last equality was obtained from the definition of the adjoint representation given in equation (\ref{adjoint_def}), along with the following identity on the generators of SU(2) \cite{Haber2021}
\begin{equation}
(\hat{T}^{a})_{ij} (\hat{T}^{a})_{kl}=\frac{1}{2}\left( \delta_{il}\delta_{jk}-\frac{1}{2}\delta_{ij}\delta_{kl}\right).
\end{equation}
Finally, combining all the results above we can write the transformation of the left electric field as
\begin{equation}
\hat{\Theta}\hat{L}_{n}^{a}\hat{\Theta}^{\dagger}=\hat{L}_{n}^{a}-\sum_{m>n} \left( \hat{U}_{n-1}^{\text{adj}}\hat{U}_{n-2}^{\text{adj}}\cdots \hat{U}_{1}^{\text{adj}}\right)_{ab}\hat{Q}_{m}^{b}. \label{leftelec_transformation}
\end{equation}
Note that since the first site has no charges to its left, the left electric field on the first site transforms as
\begin{equation}
\hat{\Theta} \hat{L}_{1}^{a} \hat{\Theta}^{\dagger}=\hat{L}_{1}^{a}-\sum_{m>1}\hat{Q}_{m}^{a}. \label{Q1thetatransform}
\end{equation}
The transformation rule of the right electric field can easily be obtained from the relation $\hat{R}_{n}^{a}=(\hat{U}_{n}^{\text{adj}})_{ab}\hat{L}_{n}^{b}$ and equation (\ref{leftelec_transformation}), one finds 
\begin{equation}
\hat{\Theta}\hat{R}_{n}^{a}\hat{\Theta}^{\dagger}=\hat{R}_{n}^{a}-\sum_{m>n} \left( \hat{U}_{n}^{\text{adj}}\hat{U}_{n-1}^{\text{adj}}\cdots \hat{U}_{1}^{\text{adj}}\right)_{ab}\hat{Q}_{m}^{b}. \label{rightelec_transformation}
\end{equation}
From equation (\ref{QtransformationbyW}), we derive the transformation of the non-Abelian charges as
\begin{equation}
\hat{\Theta}\hat{Q}_{n}^{a}\hat{\Theta}^{\dagger}=\left( \hat{U}_{n-1}^{\text{adj}}\hat{U}_{n-2}^{\text{adj}}\cdots \hat{U}_{1}^{\text{adj}}\right)_{ab}\hat{Q}_{n}^{b} .
\label{Qthetatransform}
\end{equation}
Gauss's law therefore transforms as
\begin{equation}
\hat{\Theta} \hat{G}_{n}^{a}\hat{\Theta}^{\dagger}=\hat{L}_{n}^{a}-\hat{R}_{n-1}^{a}=0, \, \, \forall n>1. 
\end{equation}
In the rotated frame, Gauss's law can thus be solved recursively and gives 
\begin{equation}
\hat{L}_{n}^{a}=(\hat{U}_{n-1}^{\mathrm{adj}}\hat{U}_{n-2}^{\mathrm{adj}}\cdots \hat{U}_{1}^{\mathrm{adj}})_{ab} \hat{L}_{1}^{b}, \label{integratedGauss}
\end{equation}
where we have used the relation between the right and left electric field via the adjoint representation. As previously pointed out, the transformation for the first site of the chain has to be studied on its own. For $n=1$, Gauss's law transforms as $\hat{\Theta} \hat{L}_{1}^{a} \hat{\Theta}^{\dagger}=\hat{R}_{0}^{a}+\hat{Q}_{1}^{a}$, where we have used the fact that the right hand side is invariant under the $\hat{\Theta}$ transformation. In fact, $\hat{R}_{0}^{a}$ is invariant, since $\hat{\Theta}$ is independent of $\hat{\Omega}_{0}^{a}$ by construction, furthermore $\hat{Q}_{1}^{a}$ does not have any charge to its left, and is thus invariant from equation (\ref{Qthetatransform}).  We now use equation (\ref{Q1thetatransform}) to write the left electric field on the first site as $\hat{L}_{1}^{a}=\hat{R}_{0}^{a}+\sum_{m\geq 1}\hat{Q}_{m}^{a}$. Inserting this expression into the integrated version of Gauss's law (\ref{integratedGauss}), we obtain 
\begin{equation}
\hat{L}_{n}^{a}=(\hat{U}_{n-1}^{\mathrm{adj}}\hat{U}_{n-2}^{\mathrm{adj}}\cdots \hat{U}_{1}^{\mathrm{adj}})_{ab} \left( \hat{R}_{0}^{b}+\sum_{m\geq 1}\hat{Q}_{m}^{b} \right), \forall n\geq 1,
\end{equation}
which expresses the left electric field as a product of orthogonal (operator valued) matrices with the total non-Abelian charges. Finally, by inserting this expression on the right hand side of equation (\ref{leftelec_transformation}) we obtain 

\begin{equation}
\hat{\Theta} \hat{L}_{n}^{a} \hat{\Theta}^{\dagger}= (\hat{U}_{n-1}^{\mathrm{adj}}\hat{U}_{n-2}^{\mathrm{adj}}\cdots \hat{U}_{1}^{\mathrm{adj}})_{ab} \left( \hat{R}_{0}^{b}+\sum_{m\leq n}\hat{Q}_{m}^{b} \right),
\end{equation} 
which by virtue of the orthogonal nature of the adjoint representation, leads to the transformed electric field used in equation (\ref{SIeq:rotatedelec}).

By setting the background field $\hat{\boldsymbol{R}}_{0}=0$, and
using the fact that the non-Abelian charges commute on different sites, we can rewrite the chromoelectric energy term appearing in the Hamiltonian as
\begin{equation}
\hat{H}_{e}=\sum_{n=1}^{N-1}(N-n)\hat{\boldsymbol{Q}}_{n}^{2}+2\sum_{n=1}^{N-2}\hat{\boldsymbol{Q}}_{n}\cdot\sum_{m=n+1}^{N-1}(N-m)\hat{\boldsymbol{Q}}_{m}, \label{chromoelectricfield}
\end{equation} 
which exhibits long-range interaction between the non-Abelian charges. The formula given above is completely general and can be used for both Abelian and non-Abelian models as long as the appropriate expressions for the charges are used. For instance, in the staggered formulation of the Abelian $\operatorname{U}(1)$ theory, the electric charge at site $n$ is given by $\hat{Q}_{n}=\hat{\phi}_{n}^{\dagger}\hat{\phi}_{n}-( 1-(-1)^{n})/2$. Substituting this expression in equation (\ref{chromoelectricfield}), we recover the electric energy of the Abelian $\operatorname{U}(1)$ Kogut-Susskind Hamiltonian which has been studied both theoretically and numerically \cite{hamer_series_1997, banuls_density_2017} and has already been implemented on a quantum computer \cite{muschik_u1_2017,martinez_real-time_2016}. 

In the rotated frame, the Kogut-Susskind Hamiltonian is exclusively represented in terms of fermionic degrees of freedom. Although the gauge fields have been eliminated, their interaction with the matter field has been directly incorporated into the Hamiltonian by virtue of Gauss's law. Furthermore, we emphasize that gauge field observables are still accessible in this approach even though they do not appear explicitly in the transformed Hamiltonian.

\subsection{Qubit encoding}
\label{SI:qubit}
\begin{figure}[t]
	\includegraphics[width=\columnwidth]{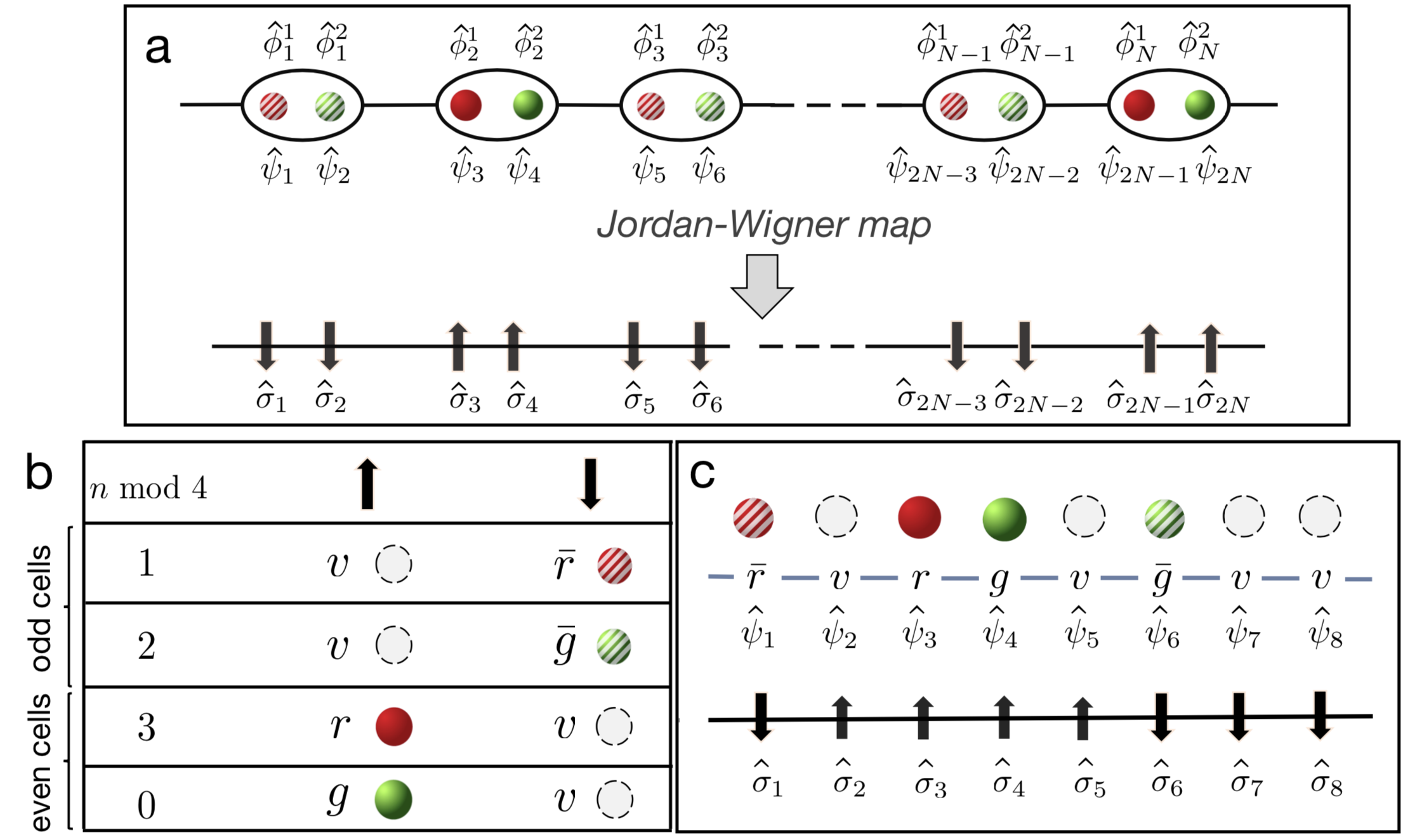} 
	\caption{\textbf{Qubit encoding of the SU(2) Kogut Susskind Hamiltonian.} \textbf{a.} Decolorization of the fermionic field by doubling of the number of lattice sites. The color degrees of freedom are encoded in a single component fermionic field $\hat{\psi}_{n}$. Even sites of the new lattice are occupied by fermions of green color while odd sites host the ones of red color. The single component fermionic fields are subsequently mapped to qubits by performing a Jordan-Wigner transformation.  \textbf{b.} Translation table between spin and fermion degrees of freedom in the staggered formulation. The matter or antimatter type of a site $n$ is determined by the first column of the table: occupied even cells (defined by $n \, \mathrm{mod} \, 4 =3,0$) translate to the presence of matter (red $r$ or green $g$) while unoccupied odd cells ($n \, \mathrm{mod} \, 4 =1,2$) translate to the presence of antimatter (antired $\bar{r}$ or antigreen $\bar{g}$). The vacuum ($v$) is represented by unoccupied even cells and occupied odd cells. \textbf{c.} For illustration, a matter configuration with four spatial spatial sites is shown in the fermion occupation number basis $\ket*{\begin{array}{c c c c} \bar{r} & r & v &v \\  v & g & \bar{g} &v  \end{array}}$ (upper panel) and after mapping the fermion fields into Pauli spin operators $\ket*{\downuparrows \upuparrows \updownarrows \downdownarrows}$ (lower panel).} \label{SIfigureencoding}
\end{figure}

In this section, we discuss the transformation from the fermionic Hamiltonian in equation (\ref{SIeq:KSham}) to a formulation that consists of qubits only. The transformation is achieved in two steps: first, the size of the lattice is doubled and the colored fermionic fields are distributed among the new lattice sites by defining the single component fields $\hat{\psi}_{2n-1}=\hat{\phi}_{n}^{1}$, $\hat{\psi}_{2n}=\hat{\phi}_{n}^{2}$, $n=1,2,\dots,N$. Note that each site now hosts one fermion with a definite color, instead of two fermions of different color.
As illustrated in Fig.~\ref{SIfigureencoding}a, odd (even) sites of the new lattice are occupied by fermions of red (green) color. By construction, the new field is of fermionic nature as it just corresponds to a relabelling of the existing fermionic fields. The color degree of freedom has thus been absorbed at the cost of doubling the size of the lattice. To have an easier interpretation, the new lattice can be divided in cells $(\hat{\psi}_{2n-1},\hat{\psi}_{2n})$, with $n=1,2,\dots,N$. For $n$ odd, the fields represent antimatter particles, while cells with even $n$ represent matter particles (see Fig.~\ref{SIfigureencoding}b). The parity of the sites thus encodes the color degree of freedom, while the parity of the cells determines the matter-antimatter nature of the particles on the site.

In the second step, the single component fermionic field $\hat{\psi}_{n}$ is mapped into $\frac{1}{2}$-spin operators by means of a Jordan-Wigner transformation \cite{JordanWigner}
\begin{equation}
\hat{\psi}_{n}=\prod_{l<n}\left(-\hat{\sigma}_{l}^{z}\right)\hat{\sigma}_{n}^{-}, \quad \hat{\psi}_{n}^{\dagger}=\prod_{l<n}\left(-\hat{\sigma}_{l}^{z}\right)\hat{\sigma}_{n}^{+},
\end{equation}
where $\hat{\sigma}_{n}^{\pm}=(\hat{\sigma}_{n}^{x}\pm i \hat{\sigma}_{n}^{y})/2$. The string factor $\prod_{l<n}\left(-\hat{\sigma}_{l}^{z}\right)$ permits to recover the correct fermionic anticommutation relations for the field $\hat{\psi}_{n}$.
Writing the rotated Hamiltonian in equation~(\ref{SIeqrotatedHam}) in terms of the fermionic field $\hat{\psi}_{n}$ and  applying the Jordan-Wigner transform, we find that the Kogut-Susskind Hamiltonian takes the form
\begin{equation}
\hat{H}= \tilde{m} \hat{H}_{\text{m}} + \frac{1}{x} \hat{H}_{\text{el}} + \hat{H}_{\text{kin}}, \label{SIdimensionlessHam}
\end{equation}
where we rescaled by the lattice spacing $a_l$, and added a constant to normalize the ground state energy to zero in the limit of $x\to0$. In particular, we introduced the Hamiltonian parameters $\tilde{m}=a_l m$ and coupling strength $x=\frac{1}{a_l^{2}g^{2}}$.
The explicit form of the kinetic term is given by 
\begin{equation}
\hat{H}_{\text{kin}} =-\frac{1}{2}\sum_{n=1}^{N-1}\left( \hat{\sigma}_{2n-1}^{+}\hat{\sigma}_{2n}^{z}\hat{\sigma}_{2n+1}^{-}+\hat{\sigma}_{2n}^{+}\hat{\sigma}_{2n+1}^{z}\hat{\sigma}_{2n+2}^{-}+\mathrm{H.C.}\right), \label{kineticterm}
\end{equation}
the mass term reads
\begin{equation}
\hat{H}_{\text{m}}=\frac{1}{2}\sum_{n=1}^{N}(-1)^{n}\left( \hat{\sigma}_{2n-1}^{z}+\hat{\sigma}_{2n}^{z}\right)+ N, \label{massHamiltonian}
\end{equation}
and finally, the chromoelectric Hamiltonian is expressed as
\begin{align}
\notag	\hat{H}_{\text{el}}&=\frac{3}{16}\sum_{n=1}^{N-1}(N-n)(1-\hat{\sigma}_{2n-1}^{z}\hat{\sigma}_{2n}^{z})\\ \notag &+\frac{1}{16}\sum_{n=1}^{N-2}\sum_{m>n}^{N-1}(N-m)\left(\hat{\sigma}_{2n-1}^{z}-\hat{\sigma}_{2n}^{z}\right)\left(\hat{\sigma}_{2m-1}^{z}-\hat{\sigma}_{2m}^{z}\right)\\
&+\frac{1}{2}\sum_{n=1}^{N-2}\sum_{m>n}^{N-1}(N-m)\left(\hat{\sigma}_{2n-1}^{+}\hat{\sigma}_{2n}^{-}\hat{\sigma}_{2m}^{+}\hat{\sigma}_{2m-1}^{-}+\mathrm{H.C.}\right). \label{SIqubitelectric}
\end{align}

The gauge fields do no longer appear explicitly in the new formulation at the expense of introducing long range spin-spin interactions that are present in the chromoelectric Hamiltonian. Crucially, some of these terms are off-diagonal interactions, and are a direct consequence of the non-Abelian nature of the model, and do not appear for instance in the U(1) Abelian Schwinger model. 
The qubit formulation of the SU(2) Kogut-Susskind Hamiltonian opens the way to the implementation of a non-Abelian model containing both matter and gauge fields on a quantum computer.

\subsection{Symmetries and eigenstates}
\label{SI:singletstates}
In this section we discuss the structure of the eigenstates of the qubit Hamiltonian given in \eqref{SIdimensionlessHam}, considering the symmetry constraints of the theory. 

Since we consider the neutral charge sector, we can expand the eigenstates of the Hamiltonian in the basis of the zero mode eigenstates of the total non-Abelian charges, which in the qubit formulation read
\begin{align}
\hat{Q}_{\mathrm{tot}}^{x}&=\frac{1}{2}\sum_{n=1}^{N}\left( \hat{\sigma}_{2n-1}^{+}\hat{\sigma}_{2n}^{-}+\mathrm{H.C}\right), \label{Qtotx_qubitform}\\
\hat{Q}_{\mathrm{tot}}^{y}&=\frac{i}{2}\sum_{n=1}^{N}\left( \hat{\sigma}_{2n-1}^{-}\hat{\sigma}_{2n}^{+}-\mathrm{H.C}\right), \label{Qtoty_qubitform}\\
\hat{Q}_{\mathrm{tot}}^{z}&=\frac{1}{4}\sum_{n=1}^{N}\left( \hat{\sigma}_{2n-1}^{z}-\hat{\sigma}_{2n}^{z}\right),
\end{align}

while the baryon number is given by
\begin{equation}
\hat{B} = \frac{1}{4} \sum_{n=1}^{2N} \hat{\sigma}_n^z.
\end{equation}

The $z$-component of the total non-Abelian charge is diagonal, therefore it is easy to find the eigenstates with eigenvalue zero. We note that the action of $\hat{Q}_{\mathrm{tot}}^{z}$ on a cell with spins pointing in the same direction gives zero, and the only non-zero contribution comes from cells with antiparallel spins. In particular, $\hat{Q}_{\mathrm{tot}}^{z}\ket{\updownarrows}_{k}=\frac{1}{2}\ket{\updownarrows}_{k}$ and $\hat{Q}_{\mathrm{tot}}^{z}\ket{\downuparrows}_{k}=-\frac{1}{2}\ket{\downuparrows}_{k}$, where $\ket{\updownarrows}_{k}$ corresponds to the spin configuration at the spatial site (or cell index)  $k$. The first qubit in the ket $\ket{\updownarrows}_{k}$ is thus at position $2k-1$ and the second one at position $2k$ of the encoded lattice. As a consequence, in order to be an eigenstate of $\hat{Q}_{\mathrm{tot}}^{z}$ with eigenvalue zero, a basis state must contain as many cells of type $\ket{\updownarrows}$ as $\ket{ \downuparrows}$. If we call $n_{\downuparrows}$ the number of such cells appearing in the basis state and $n_{\downdownarrows}$ the number of cells with both spins pointing down, then it is easy to see that the baryon quantum number of such state is $B=N/2-n_{\downdownarrows}-n_{\downuparrows}$ with $n_{\downuparrows}=0,1,\dots,N/2$ and $n_{\downdownarrows}=0,1,\dots, N-2n_{\downuparrows}$. This clearly shows that the baryon quantum number of a physical state is an integer $B=-N/2,-N/2+1\dots,N/2$. 

Among all the eigenstates of $\hat{Q}_{\mathrm{tot}}^{z}$ with eigenvalue zero, some of them must be combined in order to be annihilated by the two other non-Abelian charges. This is obvious from the expression (\ref{Qtotx_qubitform}) and (\ref{Qtoty_qubitform}), where we see that these operators induce non-diagonal transitions between the computational basis states. As a simple illustration, let us consider the state $\ket{\updownarrows \downuparrows \upuparrows \upuparrows}$. It contains one cell of type $\ket{\updownarrows}$ and one cell $\ket{\downuparrows}$ and is thus an eigenstate of $\hat{Q}_{\mathrm{tot}}^{z}$ with eigenvalue zero as per our discussion above. It is however easy to see that this state alone is not an eigenstate of the two other non-Abelian charges $\hat{Q}_{\mathrm{tot}}^{x,y}$. In order to be a simultaneous eigenstate of the three non-Abelian charges, it must be combined with its companion state in the table in the following way $\left(\ket{\updownarrows \downuparrows \upuparrows \upuparrows} -\ket{\downuparrows \updownarrows \upuparrows \upuparrows}\right)/\sqrt{2}$. 
For an arbitrary number of sites $N$, the construction discussed above is a non-trivial task. This is why the design of a circuit incorporating the neutral charge constraint is generally harder and one can rely on the variational algorithm to restore the right color symmetry. This approach was for instance used to obtain our results in the $N=6$ case (see Methods). For small lattice sizes, it is however possible to impose directly the color symmetry into the design of the circuit, as we did in Methods for the circuit generating the color symmetric ansatz for $N=4$ spatial sites in the sector with baryon number $B=1$.

\end{document}